%% file: 3536_ms.tex
\documentclass{aa} 
\usepackage{graphicx}
\usepackage{txfonts}
\newcommand{\etal}{{\em et al.}}

\begin{document}


\title{Astrochemistry of Sub-Millimeter Sources in Orion} 
\subtitle{Studying the Variations of
Molecular Tracers with Changing Physical Conditions}

\author{
Doug Johnstone\inst{1,2} \and
Annemieke M. S. Boonman\inst{3} \and
Ewine F. van Dishoeck\inst{3}
}
\offprints{D. Johnstone}

\institute{
National Research Council Canada, 
Herzberg Institute of Astrophysics, 
5071 West Saanich Road, Victoria, B.C., V9E 2E7, Canada\\
\email{doug.johnstone@nrc-cnrc.gc.ca}
\and
Department of Physics \& Astronomy, University of Victoria,
Victoria, BC, V8P 1A1, Canada
\and
Leiden Observatory, P.O. Box 9513, 2300 RA Leiden, The
Netherlands
}

\date{}


\abstract{
Cornerstone molecules (CO, H$_2$CO, CH$_3$OH, HCN, HNC, CN, CS, SO)
were observed toward seven sub-millimeter bright sources in the Orion 
molecular cloud in order to quantify the range of conditions for which 
individual molecular line tracers provide physical and chemical information. 
Five of the sources observed were protostellar, ranging in 
energetics from $1 - 500\,L_\odot$, while the other two sources were 
located at a shock front and within a photodissociation region (PDR).

Statistical equilibrium calculations were used to deduce from
the measured line strengths the physical conditions within
each source and the abundance of each molecule. 
In all cases except the shock and the PDR, the abundance of CO with
respect to H$_2$ appears significantly below (factor of ten) the 
general molecular cloud value of $10^{-4}$.  
{Formaldehyde measurements were 
used to estimate a mean temperature and 
density for the gas in each source. Evidence was found for trends between
the derived abundance of CO, H$_2$CO, CH$_3$OH, and CS
and the energetics of the source, with hotter sources having higher 
abundances.} Determining whether this is due to a linear progression
of abundance with temperature or sharp jumps at particular temperatures
will require more detailed modeling.
The observed methanol transitions require high temperatures ($T>50$
K), and thus energetic sources, within all but one of the observed
protostellar sources. The same conclusion is obtained from
observations of the CS 7-6 transition. Analysis of the HCN and HNC 4-3
transitions provides further support for high densities $n> 10^7$
cm$^{-3}$ in all the protostellar sources.

The shape of the CO 3--2 line profile provides evidence for internal 
energetic events (outflows) in all but one of the protostellar sources, and
shows an extreme kinematic signature in the shock region. In general,
the CO line and its isotopes do not significantly contaminate the
850$\,\mu m$ broadband flux (less than 10\%); however, in the shock region
the CO lines alone account for more than two thirds of the measured
sub-millimeter flux.  In the energetic sources, the combined flux from
all other measured molecular lines provides up to an additional
few percent of line contamination.
}

\maketitle


\section{INTRODUCTION}
\label{sec:intro}

The study of structure within the star-forming regions of molecular
clouds has benefitted significantly from observations of
many molecular lines, each tracing specific chemical and physical
conditions. Most studies either focus on surveys using
particular tracers, such as carbon monoxide, N$_2$H$^+$, or ammonia, 
in order to map the column density and kinematics of the gas 
(Myers 1999), or have 
focused on individual sources, producing detailed molecular catalogues in 
several to dozens of species designed to constrain the physical and chemical 
morphology of the region (van Dishoeck \& Blake 1998).

Several molecules have emerged as particularly good diagnostics of the
conditions and chemistry near young stellar objects. For example,
H$_2$CO has many lines that are readily observed at sub-millimeter
wavelengths, and whose ratios are either good temperature or density
tracers (e.g., Hurt, Barsony, \& Wooten 1996, Mangum, Wooten, \& Barsony 1999, 
Mitchell \etal\ 2001).  Analysis of dust continuum and line emission
has shown that temperature and density gradients exist across the
protostellar envelopes, with temperatures varying from the inner to
the outer region from $>100$~K to less than 20~K and densities from
$>10^7$ cm$^{-3}$ to $\sim 10^4$ cm$^{-3}$ (e.g., van der Tak \etal\
2000a, Shirley, Evans, \& Rawlings 2002, J{\o}rgensen, Sch\"oier, \& 
van Dishoeck 2002). The chemistry
responds to these changes. Molecules freeze-out onto the grains in the
cold outer parts of the clouds, where they can form new species
through grain-surface reactions. The composition of icy mantles can be
determined through infrared absorption, with species such as H$_2$O,
CO, CO$_2$ and CH$_3$OH known to have high ice abundances (e.g., Gibb
\etal\ 2000). In the inner region close to the protostar, the grains
are heated and the ices are observed to evaporate back into the gas
(e.g., van Dishoeck \& Helmich 1996; Boonman \& van Dishoeck 2003). 
High temperature gas-phase
reactions between evaporated species can subsequently lead to high
abundances of complex organic molecules observed in high-mass hot
cores (e.g., Rodgers \& Charnley 2001). Depending on the evolutionary
state of the source, different chemical characteristics become more
prominent.

In this paper we consider a selection of cornerstone molecules and
study a variety of independent locations within the 
Integral Shaped Filament (ISF)
in Orion A, along with NGC 2071IR in Orion B, 
in order to quantify the range of
conditions for which individual molecular line tracers provide
physical or chemical information. This is primarily a morphological study, 
comparing the differences in molecular line emission across sources of
varying physical conditions using simplified equilibrium modeling,
in order to search for obvious diagnostic features within
the data. The list of sources observed
in Orion A was compiled from a sensitive dust continuum survey 
of the ISF (Johnstone \& Bally 1999) and includes both highly enshrouded
sub-millimeter bright sources, possibly protostellar, through more
evolved protostars, a bright PDR knot, and a shock front. Many of 
the young stellar objects are of intermediate mass and thus the
observations bracket sub-millimeter studies of high-mass 
(e.g. Hatchell \etal\ 1998; van der Tak \etal\ 2000a) and low-mass 
(e.g. Blake \etal\ 1994, 1995; van Dishoeck \etal\ 1995; 
Buckle \& Fuller 2002; J\o rgensen \etal\ 2002) 
star-forming regions. The
selection of cornerstone molecules and molecular transitions was chosen
based on previous experience in detailed studies of individual
objects. 

\section{The Orion Integral Shaped Filament}
\label{sec:ISF}

The ISF (Bally \etal\  1987) in the northern
portion of the Orion A molecular cloud is the nearest site with active 
high-mass through low-mass star formation. The region is associated with the 
Orion Nebula and the Trapezium Cluster of 700 young stars (Hillenbrand 1997;
Hillenbrand \& Hartmann 1998), contains the OMC1 cloud core immediately behind
the Nebula, and two other extensively studied star-forming cores, 
OMC2 and OMC3 located about 15\arcmin\ and 25\arcmin\ to the north 
(Castets \& Langer 1994). To the south the region becomes much quieter
with OMC4 (Johnstone \& Bally 1999) showing no bright sub-millimeter sources.
The ISF has spawned several thousand young stars in the past few million years
and contains dozens (possibly hundreds) of embedded young stellar objects
(Chini \etal\ 1997, Johnstone \& Bally 1999) which power dozens of molecular 
outflows,  Herbig-Haro objects (Reipurth, Bally, \& Devine 1997), and 
molecular hydrogen emitting shocks and jets (Yu, Bally, \& Devine 1997).
The abundance of star formation, at various stages of evolution, within a
single, nearby, molecular cloud provides a unique setting for conducting
tests of astrochemical diagnostics. 

\subsection{The Individual Sources}
\label{sec:sources}

The sources chosen for this study were primarily drawn from the 450 and 
850\,$\mu$m dust emission 
maps of the ISF (Johnstone \& Bally 1999; Fig.\ \ref{f_orion}). Supplemental
information on the properties of the sources comes from measurements
at 350\,$\mu$m (Lis \etal\ 1998), 1300\,$\mu$m (Chini \etal\ 1997), 
$3.6\,$cm VLA data (Reipurth, Rodriguez, \& Chini 1999), and the IRAS 
point source catalogue. Additional 2\,$\mu$m molecular hydrogen observations
(Yu \etal\ 1997) of the ISF provide evidence of local heating expected 
within shocks, and delineate the location of molecular jets.
The seven sources were chosen to cover a wide selection of physical 
conditions in order to test the hypothesis that molecular line observations
provide sensitive diagnostics.

The coldest sub-millimeter source in our sample is MMS6, 
$T_d \sim 15\,$K as measured by the 
sub-millimeter fluxes from 350 to 1300\,$\mu$m  (Lis \etal\ 2001). 
MMS6 is also the 
brightest 850\,$\mu$m source within the OMC3 region of the ISF, with a peak 
flux of 7.4 Jy within a 15 arcsecond beam (Johnstone \& Bally 1999). 
The total mass within 
the envelope is estimated to be $36\,M_\odot$ (Chini \etal\ 1997).
There is no corresponding
IRAS point source, although the background brightness of the region makes 
point source detection difficult, suggesting that there is no strong internal 
heating source. The sub-millimeter luminosity, energy emitted at wavelengths greater
than 350$\,\mu m$, is $1.2\,L_\odot$; however,
the lack of far-infrared emission measurements provides only an upper limit on
the total bolometric luminosity of  $< 60\,L_\odot$
(Chini \etal\ 1997).
Despite the very cold appearance of MMS6 in the sub-millimeter, and
thus the possibility that it is a pre-stellar clump heated only from the
outside by the strong interstellar radiation field (ISRF) in Orion,
MMS6 has been classified as a Class 0 protostar (Chini \etal\ 1997). 
Yu \etal\ (1997) 
observed knots of H$_2$ directly north and south of MMS6 which are
interpreted as bow shocks within a molecular outflow from the continuum
source. Chini \etal\ (1997) report substantial high-velocity
red-shifted emission in $^{12}$CO centered on MMS6, although no
corresponding blue-shifted lobe was observed. More recent observations
by Aso \etal\ (2000) associate the CO outflow with a different sub-millimeter
source. VLA observations at
3.6$\,$cm (Reipurth \etal\ 1999) locate a radio source coincident with
MMS6. Such coincidences are common among Class 0 sources and have been 
interpreted as thermal Bremsstrahlung  radiation produced in shocks 
within the inner part of the molecular outflow. Thus, while MMS6 is 
possibly a pre-collapse clump, it is likely that the sub-millimeter source 
enshrouds a very young Class 0 protostar.

Despite being three times fainter than MMS6 at 850\,$\mu$m, MMS9 appears
to be the source of the most powerful outflow in the OMC2/3 region 
(Yu \etal\ 2000). The spectral index for MMS9 is less steep than observed
in MMS6, suggesting a warmer temperature in the dust envelope, 
$T_d \sim 20\,$K (Chini \etal\ 1997).  The total envelope mass is
estimated to be $10\,M_\odot$, considerably lower than the envelope mass
of MMS6. The IRAS flux limits constrain the luminosity of the embedded
source to be $< 94\, L_\odot$ (Chini \etal\ 1997). Despite
having no coincident IRAS point source, MMS9 is expected to contain a 
well-advanced Class 0 protostar. Reipurth \etal\ (1999) observed
3.6 cm continuum emission at this location.
The rising infrared background toward 
MMS9 may be responsible for the lack of an IRAS point source detection.

The only IRAS point source within the OMC3 region is IRAS 05329-0505,
coincident with MMS7. From the wavelength of the peak emission, 
$\lambda_{\rm max} \sim 60$\,$\mu$m, the warm component of the dust
can be estimated, $T_d \sim 50\,$K. In the sub-millimeter MMS7 appears 
quite similar to
MMS9, although the envelope temperature is somewhat higher 
$T_d \sim 26\,$K (Chini \etal\ 1997). The total envelope mass is 
$\sim 8 M_\odot$, comparable to the envelope around MMS9, and the
bolometric luminosity is $\sim 76\,L_\odot$.
Reipurth \etal\ (1999) observed 3.6 cm continuum emission at this location,
providing further evidence for an embedded Class 0 source.
MMS7 appears to be in a somewhat more advanced stage of star 
formation, powering a giant bipolar Herbig-Haro flow (Reipurth \etal\ 1997)
and harboring both a central FU Ori-like source (Reipurth \& Aspin 1997)
and an optical reflection nebula (Wolstencroft \etal\ 1986).

Further south along the ridge, near the center of the OMC2 molecular core,
lies the far infrared source FIR4 (Mezger, Zylka, \& Wink 1990), 
also observed as an IRAS point source.  The region around FIR4 harbors
a cluster of near-infrared sources (Johnson \etal\ 1990) although none
are explicitly coincident with the far-infrared source.  
FIR4 is the brightest sub-millimeter source in the OMC2 region and contains 
a VLA radio source (Reipurth \etal\ 1999), leading to identification as a
Class 0 object. Its proximity to the near IR cluster suggests, however, 
that it might be a more evolved envelope. The presence of
far-infrared  emission provides evidence for warm dust $T_d \sim 40\,$K 
(Mezger \etal\ 1990), although the sub-millimeter emission traces the much 
colder outer envelope $T_d \sim 15\,$K. The total envelope mass is 
$\sim  34\,M_\odot$ and the luminosity is 
$\sim 400\, L_\odot$ (Mezger \etal\ 1990).

The final protostellar source in our sample is NGC2071IR in the Orion B
North molecular cloud (Johnstone \etal\ 2001; Motte \etal\ 2001; Harvey \etal\
1979). Significant far-infrared emission from the source (Butner \etal\ 1990) 
provides evidence of an internal source and a warm dust $T_d > 50\,$K component
providing a reprocessed luminosity $\sim 500\, L_\odot$.
The total envelope mass, derived from the sub-millimeter
dust emission, is $12 - 30\,M_\odot$ (Johnstone \etal\ 2001). 
NGC2071 contains one of the most powerful outflows known
(Bally \& Lada 1983).

Along with the five protostellar candidates, two sub-millimeter bright
regions which are not associated with individual protostars were
also observed. A sub-millimeter bright location, PDR1, along the warm photodissociation region
(the Orion Bar) separating the HII region M42 (the Orion Nebula)  
from the Orion A 
molecular cloud was included to monitor changes in the molecular line 
diagnostics in regions of enhanced UV fluxes and higher excitation 
temperatures (see also Hogerheijde, Jansen, \& van Dishoeck 1995). 
Also, a strong shock front, SK1, located via bright
2\,$\mu$m H$_2$ emission (Yu \etal\ 1997), was included to test for
shock diagnostics. The shocked region is also visible in the spectral
index map of the ISF (Johnstone \& Bally 1999) as a region with anomalously
shallow slope, requiring either a {\it lower} temperature than the 
surrounding material, a shallower dust emissivity profile, or
contamination of the 850\,$\mu$m continuum measurement by molecular lines.

A summary of the physical properties of the seven sources is presented in 
Table \ref{t_prop}.  The sources are ordered from the coldest, least energetic,
to the warmest, most energetic, with the PDR and shock front noted
separately. 

\section{Observations}

\subsection{Choice of Molecules}

Eight different molecules and their isotopes have been observed in
order to find useful diagnostics to study physical and chemical
differences between the sources in our sample. Among these are
carbon-, oxygen-, nitrogen-, and sulphur-bearing species, allowing
for a study of differentiation in the chemical abundances.
Abundant CO is observed toward all sources.
Nevertheless, it is a useful molecule to study depletion effects,
which are expected to be largest for the coldest, less evolved sources.
In those sources where CO is not too depleted, it quickly becomes optically
thick. In that case, the optically thin isotope C$^{17}$O can be used
to derive H$_2$ column densities.

Formaldehyde, H$_2$CO, is a slightly asymmetric rotor molecule, providing
it with many of the temperature and density diagnostic features of purely
symmetric rotors. Mangum \& Wootten (1993) and van Dishoeck \etal\ (1993)
list many of the strengths of
this molecule as a probe of the physical conditions in dense molecular clouds.
Methanol, CH$_3$OH, is a complex molecule with many transitions in the 
sub-millimeter.
Since these lines are concentrated in complex bands spanning relatively
small wavelength ranges, this allows for efficient observations even at
relatively high spectral resolution. Methanol is also chemically important,
since it has been detected both in the gas-phase and on grains 
(e.g. Dartois \etal\ 1999).  The high gas-phase abundances of CH$_3$OH most 
likely result from evaporation off of grains (Charnley, Tielens, \& Millar 1992; 
van der Tak, van Dishoeck, \& Caselli 2000b; Sch\"oier \etal\ 2002).
 
One of the most abundant nitrogen-bearing species is HCN. Chemical models
predict that the HCN abundances increase with temperature
(Rodgers \& Charnley 2001), thus providing a useful evolutionary indicator.
Observations of massive star-forming regions confirm this
(Lahuis \& van Dishoeck 2000), although the situation for low-mass sources
is less clear (J{\o}rgensen \etal\, in prep.).
Schilke \etal\ (1992) show that the abundance ratio of HCN and HNC depends
on the density and kinetic temperature of the gas, providing a
physical diagnostic. The chemical equilibrium of HCN, HNC, and CN also
depends on the hardness of the radiation field.
  
Finally, sulphur-bearing species have been associated with
shocks and/or outflows (e.g. Orion; Blake \etal\ 1987). However, recent
studies of sulphur-bearing species in massive star-forming regions other
than Orion cannot distinguish between an origin in the circumstellar
envelope or the outflow (Keane \etal\ 2001a; van der Tak \etal\ 2003).
Thus, it is still debatable whether sulphur-bearing species are good
diagnostics of outflows. On the other hand, CS transitions have
been used to derive detailed density structure within the molecular
envelopes of star-forming regions, making this species a useful 
indicator of physical properties (van der Tak \etal\  2000a).
The H$_2$S/SO$_2$ ratio has also been proposed as a chemical clock of 
hot core regions (Charnley 1997), although observational results are still
ambiguous (Hatchell \etal\ 1998; van der Tak \etal\ 2003).

\subsection{Sub-millimeter Observations}

The molecular line observations were taken at the 15-meter James Clerk 
Maxwell telescope (JCMT)\footnote{
The JCMT is operated by the Joint Astronomy Centre on behalf of the
Particle Physics and Astronomy Research Council of the UK, the
Netherlands Organization for Scientific Research, and the National
Research Council of Canada.} on Mauna Kea in Hawaii. 
The data were obtained during three separate observing runs: 
January 1999, September 1999, and December 2000.
Table \ref{t_cal}  provides details of the instruments and settings used during each
run, along with typical receiver conditions. For each source, Table \ref{t_spec}
lists the molecular lines observed and the results of fitting one or
two Gaussian components to the integrated line intensity, the full width
at half maximum, and the peak intensity.

The data were obtained in a beam switching mode, with a 10 arcminute
switch in right ascension. The integrated line data were 
transfered from the antenna temperature scale $T^*_A$ to the main-beam 
brightness scale $T_{\rm mb}$
by dividing by the main-beam efficiency $\eta_{\rm mb}$. For the A band
receiver (210--270 GHz) the measured efficiency is  $\eta_{\rm mb} = 0.69$, 
while for the B band receiver (330--370 GHz) $\eta_{\rm mb} = 0.60$. 
The relevant
beam sizes for the JCMT are 21 and 14 arcseconds at respectively 230 and 345 
GHz. The velocity resolution ranged from $0.2-0.5\,$km\,s$^{-1}$. 
The rms noise after a typical integration time of twenty minutes was $0.30\,$K per
channel for A band observations and $0.15\,$K per channel for B band observations. 
Further details on the JCMT and the heterodyne receivers can be found on the
JCMT homepage\footnote{ {\tt http://www.jach.hawaii.edu/JACpublic/JCMT}}.

\subsection{Representative Line Profiles}

Many important differences between the sources are readily 
observed simply by looking at the spectra. The extremely strong,
and almost always self-absorbed, CO 3--2 transition reveals
information on the gas kinematics within each source
(Fig.\ \ref{f_cop}). Single component Gaussian fits to the
profiles are only possible for MMS6 (line width 4\,km\,s$^{-1}$) and 
PDR1 (3\,km\,s$^{-1}$). MMS9 requires both a narrow
component (6\,km\,s$^{-1}$), providing two thirds of the integrated
intensity, and a broad component (15\,km\,s$^{-1}$).  FIR4 appears
to be a scaled-up version of MMS9, with a narrow component
(9\,km\,s$^{-1}$) containing three quarters of the integrated intensity
and a broad component (21\,km\,s$^{-1}$). Both the shock region
and NGC2071 are poorly fit by Gaussian components and are
dominated by the extended wings.

Eight of the methanol lines are observed in a single spectral
setting and can be plotted for each source 
(see examples in Fig.\ \ref{f_ch3ohp}).
{These spectra also contain the sole observed SO
transition. In agreement with the results found by van der Tak \etal\ (2003) 
there is an apparent lack of correlation between
the strength of the SO line and the methanol lines.} As
well, differences in the relative strengths of the methanol
lines are observable between MMS6 and FIR4, indicating 
the more extreme temperature and density conditions in
the energetic infrared source. Finally, the FIR4 methanol
spectra show evidence for two components, narrow and
broad, similar to the CO spectra.

The HCN and HNC 4--3 profiles are also observed in a single
spectral setting (Fig.\ \ref{f_hcnp}). A range of
relative line strengths is seen in the spectra; however,
optical depth considerations make simple abundance determinations
more complicated (see Sect.\ \ref{sec:HCN}). The
substantial optical depth of the transitions allows for 
more detailed determinations of the source kinematics.
In all sources except PDR1,
the HCN line profile is broader than the HNC line profile
or has evidence for extended line wings.

\section{Molecular Line Survey}
\label{sec:molecules}

\subsection{Statistical Equilibrium Calculations}

The analysis of the data is done with statistical equilibrium
calculations, using the numerical code {\sc RADEX}\footnote{This program 
was originally  written by J.H. Black. Here a more recent version by D.J.\
Jansen is used.}.
The adopted method is described in detail by Jansen, van Dishoeck, \& Blake (1994),
van Dishoeck \etal\ (1993), and Jansen \etal\ (1995).
This method assumes that the source is a homogeneous spherical
region.  The decoupling of radiative transfer and level populations is done 
with the escape probability method. The optical depth $\tau$ at the line 
center is proportional to the ratio of the total column density and the line 
width of the molecule. The optical depth and the line intensity
depend only on this ratio. All variables can be calculated, using a 
geometrical assumption that they are constant throughout the region.
The radiative excitation of rotational lines in the sub-millimeter region is 
usually dominated by the 2.73 K cosmic background radiation, which peaks at
1.871\,mm.  Therefore, this external radiation field is included in the
calculations. No central radiation field is included nor infrared pumping by 
dust. Since the sources have rather low dust temperatures, this latter effect 
is expected to be negligible. 

The kinetic temperature $T_{\rm kin}$ and the density $n(\rm H_2)$ for
the homogeneous region can be estimated from line ratio calculations
where available.
The column density of molecule $X$, $N(X)$, can then be estimated using
the temperature and the density as found from line ratio calculations, such
that the line width, integrated line strength, and main beam temperature of
the emission lines can be fitted.

{Due to the simplistic treatment of the envelope structure the derived 
abundances are only roughly determined.  Rising temperature and density
gradients toward the envelope centers are expected due to heating by
the central protostellar sources 
and these conditions can influence the observed line strengths significantly.  
Additionally, these Orion sources reside within an enhanced external radiation 
field, due to the proximity of many nearby massive stars
(Li, Goldsmith, \& Menton 2003). Thus, the outer envelopes are kept 
warm, $T_d \sim 30\,$K (Johnstone \& Bally 1999). 
Modeling of such envelopes (Li \etal\ 2003, J\o rgensen \etal\, in prep.) reveal
an inward {\it decreasing} temperature gradient near the surface, encompassing
much of the mass of the envelope, coupled to an inward {\it rising} temperature 
gradient once the central heating source dominates. 

The large, externally warmed, envelopes surrounding the protostellar 
sources in Orion should allow for reasonable abundance estimates from 
the observed line intensities
using statistical equilibrium calculations, except for those transitions requiring 
either very high densities $n > 10^7\,$cm$^{-3}$ or high temperatures $ T > 50\,$K.
As a guide to the accuracy of the calculations presented in this paper, detailed 
modeling of the protostellar source IRAS 16293-2422, including careful derivation of the 
physical structure, has shown that abundance estimates from this simpler procedure are 
typically only accurate to within a factor of two or three (Sch\"oier \etal\ 2002). 
}

\subsection{Isotopes of Carbon Monoxide}

\label{sec:CO}

Within star-forming regions of molecular clouds, comparison
between the line strengths of low-lying CO transitions and the
intensity of the sub-millimeter dust continuum often suggest that
there is a depletion of CO with respect to the dust, and by
extension the molecular hydrogen abundance (see for example
Mitchell \etal\ 2001, J\o rgensen \etal\ 2002, 
Bergin \etal\ 2002, Bacmann \etal\ 2002).  The most common
explanation for this observational discrepancy between CO line 
strength and dust continuum emission is that the CO has depleted 
through freeze-out onto the grains.  The time scale for depletion
scales inversely with the density of the region, and is shorter than
$10^5$ yr for densities above a few $\times 10^4$ cm$^{-3}$.  Recent
laboratory experiments on CO and H$_2$O ices show that CO will remain
completely frozen out if the grain temperatures stay below
$\sim$20~K. Above 20~K, CO starts to evaporate back into the gas phase
in a complicated pattern depending on the ice structure, with a
fraction of the CO remaining trapped in the H$_2$O ice matrix until
the entire mantle evaporates at $\sim$100~K (Collings \etal\ 2003,
Fraser \etal\ 2001).

\subsubsection{Line strengths}

Table \ref{t_co} details the observed line strengths of CO, $^{13}$CO, and 
C$^{17}$O 3--2 transitions obtained at the JCMT, along with an estimate 
of the column of carbon monoxide along the line-of-sight. Using RADEX, 
the column is deduced from the integrated line strength of the C$^{17}$O 3--2 
line, assuming that the $^{13}$CO peak temperature provides a reasonable
estimate of the kinetic temperature of the gas.
The RADEX calculations for CO include energy levels up to $J=25$ and use 
collisional rate coefficients by Schinke \etal\ (1985). The results are
not significantly changed if the new rate coefficients by Flower (2001)
are adopted (see J{\o}rgensen \etal\ 2002).  For NGC2071, where the 
$^{13}$CO line was not observed, the temperature was assumed to be 30$\,$K,
similar to the temperature of FIR4.

The cosmic abundance ratio for isotopes of Oxygen, 
[$^{16}$O]/[$^{17}$O]$ \sim 1800$ (Wilson \& Rood 1994), was used to convert
the calculated C$^{17}$O column to an equivalent column of CO. For the 
expected conditions within the sub-millimeter clumps, 
$n > 10^{4}\,$cm$^{-3}$, there is only a slight density 
dependence on the calculated column of the gas. In most cases the CO 
peak temperature was larger than the $^{13}$CO temperature; however, the 
optical thickness of CO is extreme and hence the line profile may be dominated 
by a small fraction of hot gas associated with each source. 
There is still uncertainty comparing the thicker $^{13}$CO with the thinner 
C$^{17}$O. We estimate the reliability of the computed physical parameters by 
comparing the RADEX computed optical depth of the $^{13}$CO line, 
$\tau_{\rm mod}$, with the observationally determined optical depth 
$\tau_{\rm obs}$, measured by 
comparing the peak temperature of the C$^{17}$O against the $^{13}$CO peak 
temperature and assuming [$^{13}$CO]/[C$^{17}$O]$ = 25$ (Wilson \& Rood 1994). 
This calculation confirms that the $^{13}$CO line is optically thick in all 
cases ($\tau_{^{13}{\rm CO}} > 1.3$).
While there are deviations between the two determinations of the optical depth,
the order unity dispersion provides an equivalent estimate of the reliability 
of the measured column density of CO. 

\subsubsection{Determination of CO abundance}

The column density of molecular hydrogen within each envelope
is best determined from the sub-millimeter flux measurements. Assuming
that the 850\,$\mu$m emission observed by Johnstone \& Bally (1999) and
Johnstone \etal\ (2001) traces the bulk of the envelope and
that the dust emission is optically thin, the mean column density through 
the center of each clump is related to the peak brightness:
\begin{equation}
N(\rm{H}_2) = { S_{850}\Omega_{\rm bm} \over B_{850}(T_d) \kappa_{850}\,\mu\,m_{\rm H_2}},
\end{equation}
where $\Omega_{\rm bm}$ is the effective solid angle of the SCUBA beam (in
steradians), $B_{850}(T_d)$ is the Planck function at 850\,$\mu$m for dust radiating
at $T_d$, $\kappa_{850}$ is the dust emissivity at 850\,$\mu$m (assumed
here to be $\kappa_{850} = 0.02\,$cm${^{2}}\,$g$^{-1}$), $m_{\rm H_2}$ is the 
molecular hydrogen mass, and $\mu$ is the mean molecular weight 
(in units of $m_{\rm H_2}$). 

As shown in Appendix A, significant CO line flux can
contaminate the broadband continuum measurements; therefore, it is important 
to correct the continuum measurements for the presence of line emission.
Having both line and continuum measurements for each of the sources
in Orion allows for a direct measure of the level of this
line contamination. At 850\,$\mu$m, $1\,$Jy within the JCMT $14\,$arcsecond 
beam is equivalent to a brightness temperature of $51\,$mK. However,
the broad SCUBA 850\,$\mu$m bandpass, $30\,$GHz wide, dilutes the importance 
of individual spectral line features against the continuum. The equivalent 
line flux required to mimic a $1\,$Jy signal with SCUBA on the JCMT is 
$1540\,$K$\,$km\,s$^{-1}$. For a typical peak gas temperature of $\sim 25\,$K, 
the Gaussian component of the CO line would need to be $\sim6\,$km\,s$^{-1}$
wide to produce a ten percent, 0.1$\,$Jy, error in the continuum 
measurement. Thus, for strong continuum sources, $S > 2\,$Jy, line 
contamination should be minimal. Table \ref{t_cont} presents the CO line 
contamination for each of the six Orion sources.  
For completeness, the combined line contamination due to all
other observed lines lying within the 850\,$\mu$m bandpass is also shown.
With the exception of the shock knot SK1, the total line contamination is always less than 
about ten percent. Note that SK1 has a high 
peak brightness temperature for the CO 3--2 transition and
has very extended line wings, consistent with its identification as a hot shock 
front by Yu \etal\ (1997) on the basis of bright H$_2$ emission.

After accounting for line contamination, it is straightforward to compute the
column density of H$_2$, derived from the strength of the remaining
continuum emission, and the fractional abundance of CO for each source. These 
values are also presented in Table \ref{t_cont}. Within the PDR and the 
shock front, the abundance of gas-phase CO agrees with the expected 
value of $10^{-4}$ to within the accuracy of the computations (a factor 
of a few).
However, toward {\it each} of the sub-millimeter protostar regions the
CO gas-phase abundance is smaller than expected by about a factor of 5 - 10,
further evidence that almost all of the CO
has been depleted onto grain surfaces in the dense envelopes.
This result is in qualitative agreement with the more
detailed modeling results presented by
J\o rgensen \etal\ (2002), where low-mass Class 0 and pre-stellar sources were
observed to have similarly depleted CO. {Here, however, the outer
envelope is warm, $T_d \sim 30\,$K and the bulk of the
depletion likely occurs in an intermediate zone between the hot, centrally
heated deep interior and the externally heated surface layer. The higher
abundance of CO in MMS9 is likely due to the smaller envelope, and thus a
fractionally smaller intermediate zone.  Detailed modeling of externally
heated protostellar envelopes supports this hypothesis 
(J\o rgensen \etal\ in prep.).}

\subsection{Formaldehyde}
\label{sec:H2CO}

One of the advantages of H$_2$CO 
is the ability to observe, in the same side-band of a single spectrum, 
two transitions which together probe the temperature of
the gas under a large variation in density conditions. This
feature removes many of the observational uncertainties inherent in
sub-millimeter spectroscopy such as beam effects, weather conditions,
and instrumental gain.
The RADEX calculations for H$_2$CO include energy levels up to 
$J=10$ and use collisional rate coefficients by Green (1991). 

In the present study we observed three para-formaldehyde lines. Two of 
the lines were simultaneously observed in the 218 GHz atmospheric
window, $3_{03} - 2_{02}$ and $3_{22} - 2_{21}$. Because
transitions between the different $K_p$ ladders are radiatively
forbidden, the ratio of the line strengths from transitions in
different ladders depends only on the kinetic temperature of the
gas, operating as a finely tuned thermometer over an extended
density range (Mangum \& Wootten 1993). A third line was observed
at 362 GHz, $5_{05} - 4_{04}$, within the same $K_p$ ladder as the
$3_{03} - 2_{02}$ transition. Within a given $K_p$ ladder the
excitation energies do not vary dramatically and thus the ratio of
these lines provides a useful density indicator at low to
moderate densities ($n_H < 3\times 10^6\,$cm$^{-3}$) and a second
temperature probe at higher densities. The first panel in Fig.\ 
\ref{f_h2co} plots the expected 
line ratios against temperature and density.

The physical parameters derived from the formaldehyde lines
are presented in Table \ref{t_h2co}, and 
are plotted graphically for each source in Fig.\ \ref{f_h2co}. 
The spread of line ratios $R_{33}$ and kinetic temperature of the gas 
inferred from the $3_{03} - 2_{02}$ and $3_{22} - 2_{21}$ transitions
ranges from $R_{33} < 2,\  T > 300\,$K, at the location of PDR1,
to $R_{33} > 10, T < 30\,$K at the location of the sub-millimeter 
source MMS9. As expected, the intermediate mass 
star-forming regions NGC2071 and FIR4 show moderately high 
temperatures $T > 70\,$K, consistent with the observability of 
infrared dust continuum emission. Mangum \& Wooten (1993) also
observed FIR4 and NGC2071, obtaining temperatures of 85\,K and
80\,K and with a range of probable temperatures bounding
the results of this study. The only infrared source in 
OMC3, MMS7, shows no evidence for high temperatures.

The strength of the $5_{05} - 4_{04}$ transition constrains the
density of the environment from which the emission is observed.
A complication arises from the difference in the beam size at the
telescope for the $3_{03} - 2_{02}$ and $5_{05} - 4_{04}$ lines.
If the line emission fills the larger 218GHz beam, then there
are no difficulties in taking the ratio of the brightness temperatures 
of the two lines. However, if the source of the formaldehyde emission
is smaller than the 362GHz beam the ratio of the beam sizes
should be taken into account when comparing the brightness
temperatures to account for the effect of beam dilution.
Source sizes intermediate between the 218GHz and 363GHz beam
require a varying normalization.
Thus, the measured line ratio $R_{53}$, computed from the 
$5_{05} - 4_{04}$ and $3_{03} - 2_{02}$ lines, has an inherent uncertainty 
of $\sim 3$ due to the possible complications from intermediate beam dilution,
with the value given in Table \ref{t_h2co} strictly an upper limit.

For all sources the measured ratio $R_{53}$ is larger than 0.3, 
implying a high density, $n > 10^6\,$cm$^{-3}$. 
{The relatively evolved far infrared source FIR4 
shows two kinematically distinct components, a
narrow-line region ($\Delta V \sim 1.5\,$km\,s$^{-1}$) requiring an effective
density near $10^6\,$cm$^{-3}$ and broad-line component 
($\Delta V \sim 6\,$km\,s$^{-1}$) requiring a much higher density. 
While MMS6 and NGC2071 also show distinct velocity components, the 
density does not obviously vary between the components for these sources.
Intriguingly, in all three cases the temperature deduced from both the 
narrow and broad-line components are quite similar.}

With the temperature and density estimated, the column density
of para-formaldehyde can be deduced from the integrated line 
strength of the $3_{03} - 2_{02}$ transition.  The {high temperature} 
equilibrium ratio of the para to ortho forms (1:3) was used to 
convert this result into a total formaldehyde column density.
The inferred abundance of formaldehyde was then measured against 
both the H$_2$ column, deduced from the sub-millimeter dust 
continuum emission, and the measured CO column.  The results are 
presented in Table \ref{t_h2co}. The two infrared sources show 
the largest abundance of formaldehyde measured against the CO. For 
these sources the abundance ratios lie in the range [H$_2$CO]/[CO] 
$ = 2.5 - 10.0 \times 10^{-5}$. The sub-millimeter sources MMS6 and MMS9 show 
a lower abundance ratio [H$_2$CO]/[CO] $ \sim 1 \times 10^{-5}$. Disentangling
the importance of CO depletion in these data complicates any analysis
unless the formaldehyde depletes in a similar fashion to CO.
However, comparing the formaldehyde abundance to
H$_2$ produces a clear relation 
(see Table \ref {t_h2co} and Fig.\ \ref{f_abund}). 
{
The hot PDR region and the infrared sources, NGC2071 and FIR4, have the largest
formaldehyde abundance, approximately 5-10 times greater than observed for the 
sub-millimeter sources. At least half of the increase in the abundance of 
formaldehyde in the two infrared sources is due to the broad-line component.}

In low-mass Class 0 sources, typical [H$_2$CO]/[CO] abundance ratios
in the outer region are a few $\times 10^{-5}$, similar to those
observed here, but jumps in the H$_2$CO abundance by more than a
factor of 100 have been found in the inner warm envelopes, consistent
with ice evaporation (e.g., Ceccarelli \etal\ 2000, Sch\"oier \etal\
2002, Maret \etal\ in prep.). H$_2$CO is expected to be a product of
grain surface hydrogenation of CO and has been tentatively detected in
interstellar ices with an abundance of a few percent with respect to H$_2$O
ice, corresponding to $\sim 10^{-6}$ with respect to H$_2$ (Keane et
al.\ 2001b). {Thus, the trend observed here for the protostellar sources 
could be consistent with evaporation of H$_2$CO-containing ices in the warm 
interior regions, with more energetic sources containing a larger fractional 
mass of envelope at high enough temperatures to evaporate such ices.}
For high-mass sources, H$_2$CO shows a less 
clear relation with temperature (van der Tak \etal\ 2000b).

\subsection{Methanol}
\label{sec:CH3OH}

Methanol emission lines have been detected in complex bands at
241 and 338 GHz, with excitation energies ranging upward from
$T_{\rm{ex}}\sim$30~K. 
The detection of many CH$_3$OH lines in our spectra allows for
the construction of a rotation diagram for each source in order to get 
a first indication of the temperature and column density.
Described in detail by Blake \etal\ (1987), this method assumes 
that the lines are optically thin and that the excitation can be 
characterized by a single temperature. The resulting rotation diagrams, however,
are often misleading.  Direct computation of the level populations 
using RADEX shows that the higher excitation lines require increasingly
larger critical densities and thus the derived temperature from rotation
diagram fitting is heavily dependent on the density of the medium
in which the methanol resides (see also the discussion by 
Bachiller \etal\ 1995).  Figure \ref{f_ch3oh} reveals the effect on the
rotation diagram produced by increasing the critical density of
the medium while holding the kinetic temperature constant at
$100\,$K and the total column density at 
$N_{{\rm CH}_3{\rm OH}} = 10^{14}\,$cm$^{-3}$. 
Even though reasonable linear fits are produced, the
derived temperature ranges from $25\,$K for $n_{H_2} =
10^7\,$cm$^{-3}$ to $80\,$K for $n_{H_2} = 10^9\,$cm$^{-3}$.
The derived column densities also overestimate the actual amount of
methanol by factors of a few.

{
For a protostellar source with both density and temperature gradients
a proper analysis becomes even more complex. Given that higher 
excitation lines tend to have higher critical densities, it is likely 
that these lines are dominated by
emission solely from the deep interior of the envelope. Indeed, there is some
evidence that the source size decreases as the required excitation of the line 
increases (Hatchell \etal\ 1998). 
The bulk properties of the envelope derived from methanol, however, are expected 
to be only slightly weighted toward the highest density and temperature 
regions due to the small angular extent of these extreme locations.
Complicating even a detailed analysis, there is evidence for an enhanced
abundance of methanol at high temperatures, $T > 90\,$K, in protostellar
envelopes (van der Tak \etal\ 2002b, Sch\"oier \etal\ 2002). 
}

Despite the above concerns, one might hope that methanol may be 
used both as a temperature {\it and} density indicator by treating 
all of the lines through a proper statistical equilibrium calculation.
The RADEX calculations for methanol include energy levels up to 
$260\,$K and use collisional rate coefficients by Lees \& Haque (1974)
and Walmsley (private communication). 
For this paper, all the measured methanol lines were compared
with RADEX computations to produce a chi-squared fit as a function 
of temperature and density. The results are presented graphically
in Fig.\ \ref{f_chi2} and are tabulated in 
Table \ref{t_meth}.  For FIR4, the methanol measurements can 
be separated into narrow and broad line components. The narrow line 
region requires very high density and temperature $n > 10^{8}$\,cm$^{-3}$,
$T \sim 70\,$K, while the broad line component requires $n \sim
10^8$\,cm$^{-3}$, $T \sim 80\,$K. The broad line region appears to
be beam diluted, at least with respect to the larger 241GHz beam, 
requiring a normalization factor between the two sets of measurements
to produce reasonable chi-squared fits.
These results are not consistent with the formaldehyde measurements
in that higher densities are required to account for the methanol
observations.  Given a range of densities within the
surroundings of FIR4, however, the methanol observations would be most sensitive
to the highest density gas. In contrast, the other infrared source, 
NGC2071, is best fit by a moderate density, $n \sim 5 \times 10^6$\,cm$^{-3}$, 
and a moderate temperature, $T \sim 60\,$K, consistent with the
formaldehyde measurements.

The two sub-millimeter sources show some
variation in the inferred temperature. MMS6 requires a moderate
temperature $T \sim 50\,$K while MMS9 requires a low
temperature  $T < 30\,$K. For both sources, the methanol appears
to probe a dense inner region, $\sim 10^7$\,cm$^{-3}$ compared
with the formaldehyde derived envelope densities $\sim 10^6$\,cm$^{-3}$.
This is likely due to the higher critical density required to excite
the methanol transitions.  {Across all the sources there is a 
reasonable correlation between the methanol
and formaldehyde temperatures with the 
sub-millimeter sources being the coldest, the infrared sources being
moderately warm, and the PDR having the highest temperatures.}

The derived abundance of methanol is similar to that of formaldehyde.
The warm infrared sources have the highest abundance with respect
to CO. However, when compared with the total column of
hydrogen deduced from the sub-millimeter continuum measurements
(Table \ref{t_cont}),
the PDR region also shows an enhanced abundance of methanol.
{Considering that the methanol observations are predominantly 
probing higher density regions within each envelope, it is likely that
these abundance estimates are strictly lower limits. Differences
in the abundance derivations across sources may be 
affected by the extent of the warm interior within each envelope.}

Like H$_2$CO, CH$_3$OH is expected to be a product of the
hydrogenation of CO on grain surfaces (Tielens \& Charnley 1997) and
is detected in grain mantles, albeit with significant variations
between different sources (Dartois \etal\ 1999).  CH$_3$OH abundance
jumps of at least a factor of 100 have been found in various high-mass
objects (van der Tak \etal\ 2000b), whereas the well-studied low-mass
Class 0 source IRAS 16293 --2422 shows a jump by nearly three orders
of magnitude in the inner warm envelope (Sch\"oier \etal\ 2002). Such
high CH$_3$OH abundances can only be produced by evaporation from
grain mantles. Enhanced CH$_3$OH abundances have also been detected in
outflows at positions offset from the protostar, e.g., in L1157
(Bachiller \& P\'erez-Guti\'errez 1997), where the CH$_3$OH is
released from the grain mantles by shocks rather than thermal
evaporation. The general trend of an increased CH$_3$OH abundance with
temperature observed for our sources is consistent with this
picture. The presence of both narrow and broad CH$_3$OH line profiles
suggests that both evaporation mechanisms play a role in some objects
(e.g., NGC 2071).

\subsection{Nitrogen-bearing Species}
\label{sec:CN}

\subsubsection{HCN and HNC}
\label{sec:HCN}

Identical lines of both HCN and HNC 
are observed simultaneously with a single spectrograph
setting at the JCMT and thus, assuming the collisional cross sections 
for the various excitations are identical across species, the relative
abundance of the two molecules can be simply determined if the
transitions are optically thin.  To test the optical depth conditions,
the ratios of line strengths between HCN and H$^{13}$CN, and HNC
and HN$^{13}$C were computed. In all cases the HCN line was found
to be moderately to strongly optically thick ($\tau_{\rm HCN} \sim 3 - 6$)
however, only the most energetic sources FIR4 and NGC2071 were found
to have optically thick HNC lines ($\tau_{\rm HNC} \sim 2 - 3$). 
The [HCN]/[HNC] ratio tabulated in Table \ref{t_hcn} was determined by accounting 
for these optical depth effects.

For the four protostellar regions observed, the integrated [HCN]/[HNC] 
abundance ratio ranges from 5--10.  Using the chemical models by 
Schilke \etal\ (1992), this would imply low temperatures $T < 30\,$K and high
densities $n > 10^7\,$cm$^{-3}$, or high temperatures $T > 50\,$K and
moderate to low densities $n < 10^6\,$cm$^{-3}$.  For PDR1, the ratio rises 
to 70, requiring both high densities $n > 10^6\,$cm$^{-3}$ and high 
temperatures $T > 50\,$K (Schilke \etal\ 1992).  It is worth
noting that the main observational difference between the least energetic 
sub-millimeter sources, MMS6 and MMS9, and the most energetic 
sources, FIR4 and NGC2071, is in the peak brightness of the lines.
Thus, it would appear that the conditions probed by the HCN and HNC
molecules in all four sources are nearly identical but that the solid angle
covered in the most energetic sources is somewhat larger.  
RADEX calculations support this hypothesis. For the determined physical
conditions of the optically thick HCN lines the excitation temperature is 
expected to be relatively large $T_{\rm ex} > 10\,$K. The least 
energetic sources have observed peak intensities of only
$T_{\rm peak} \sim 1 \,$K, while the infrared sources show
$T_{\rm peak} \sim 15\,$K. 
Care should be taken in using chemistry to constrain temperature,
however, since the low-mass sources observed by J{\o}rgensen \etal\
(2003; in prep.) have high HCN/HNC abundance ratios at low
temperatures, which are not readily explained by the Schilke \etal\
(1992) models.

RADEX calculations for HCN, including energy levels up to $J=12$
and using collisional coefficients by Green \& Thaddeus (1974),
have been performed to derive HCN abundances using
the estimated physical conditions derived from formaldehyde. The resulting
abundances with respect to CO are $\sim$5$\times 10^{-5}$ for the
sub-millimeter sources and PDR1; however, the production of HCN appears to be 
significantly enhanced in the far infrared sources FIR4 and NGC2071. 
The HCN abundances with respect to H$_2$ vary from 
$0.6 \times 10^{-10}$ to $40 \times 10^{-10}$ (Table \ref{t_hcn}).
These latter abundances show a clear trend with increasing source
energetics. 
HCN has not been detected in interstellar ices due to a lack of clear
spectroscopic signatures, but is expected to freeze-out in cold
regions like most molecules. The HCN abundance is known to be enhanced
by orders of magnitude in the warm inner regions of massive protostars
(Lahuis \& van Dishoeck 2000; Boonman \etal\ 2001), but the high
temperature routes to HCN do not become effective until $T>200$~K
(e.g., Doty \etal\ 2002). It remains to be determined whether the
intermediate mass sources studied here have sufficient amounts of gas
at such high temperatures to enhance HCN significantly. The trends
observed here are more likely due to general evaporation of ices with
temperatures between 20 and 100~K, enhancing the overall gas-phase
abundances. 

Detailed consideration of the HCN line shows that it is usually
wider than the HNC line.  This requires that in the broad line region 
the ratio of the two molecules is much higher than the ratio of the
total integrated intensities, 
implying higher temperatures and/or higher densities if chemical
equilibrium is assumed (Schilke \etal\ 1992).  Another possible explanation
for the general lack of broad line HNC may be found in the shock model solution 
of Schilke \etal\ (1992) where HNC is preferentially destroyed within the 
interaction region, reforming downstream from the shock front. 
In particular, the broad HCN emission, 
coupled with little or no HNC emission, found in the shock knot SK1 
is likely best explained in this manner. Interferometer
data of the low-mass source IRAS 16293 --2422 also show that HNC
avoids the warm inner region, in contrast with HCN (Sch\"oier \etal\
2003; in prep.).

\subsubsection{CN}

Observations of the CN hyperfine structure at 340 GHz provides a set of 
constraints for the modeling of CN (see Table \ref{t_spec}). 
In the optically
thin limit, the ratio between the strongest and weakest hyperfine components 
should be $\sim 15$; 
however, this is only observed in the spectrum of PDR1. In all other 
regions, except possibly MMS6, the main CN component is optically thick, 
($\tau_{\rm CN} \sim 3-6$), while the weaker components
remain reasonably optically thin. Assuming that the line flux
is predominantly from the outer envelope surrounding the sources, the
column of CN derived from the optically thin hyperfine lines is
presented in Table \ref{t_hcn}.
The RADEX calculations for CN include energy levels up to $J=15$.
As found for the abundance of HCN, only the coldest, least energetic source,
MMS6 has a dramatically low abundance of CN. However, a clear correlation 
is found between the ratio [HCN]/[HNC] and [HCN]/[CN]. Assuming that all
CN is locked up in these three molecules, the total abundance of CN can
be readily determined.  Table \ref{t_hcn} displays the results of this
calculation. With the exception of the sub-millimeter source (MMS6) the derived
total abundance of CN is much more constant than any of the individual
molecular species.

Results of systematic studies of CN in low- or high-mass sources are
not yet available, but like HNC, CN seems to avoid the inner warm
region in the low-mass protostar IRAS 16293 --2422. The radical is
known to be abundant in the outer regions of PDRs (e.g., Fuente \etal\
1998), where it is produced by rapid gas-phase chemistry of CH and
C$_2$ with N. Alternatively, it can be a photodissociation product of
HCN.  Photodissociation of CN itself requires photons with energies
higher than 12 eV. {The observed lack of variation of the CN abundance
in the protostellar sources observed here implies that they are exposed to 
similar radiation fields. As mentioned earlier, the Orion region is bathed
in a significantly increased external radiation field due to the 
close proximity of massive stars. The much lower abundance within
the massive envelope of MMS6, however,
is not easily explained via this hypothesis unless the structure of
the envelope is such that the interstellar radiation field is unable
to penetrate through the surface layer and into the significant interior.}

\subsection{Sulphur-bearing Species}
\label{sect:S}

\subsubsection{CS}

Both CS and the rarer C$^{34}$S isotope were observed in the $J=5$--$4$ 
transition.
In all cases the line ratio was less than the isotopic ratio of 22 implying 
moderately optically thick conditions within the CS line but optically thin
conditions for the isotope.  Taking the formaldehyde physical conditions
for the temperature and the density, the column density of CS was determined
using RADEX (Table \ref{t_cs}). 
The RADEX calculations for CS include energy levels up to $J=12$.
The abundance with respect to H$_2$ trends with the 
source energetics, rising from a few $10^{-11}$ to a few $10^{-10}$, in the
protostellar sources.  This is a stronger trend than seen for HCN suggesting 
that the more energetic sources are providing more heating to their outer 
envelope.  The abundance rises sharply to a few $10^{-9}$, in the 
warm PDR1 source. 

For four of the sources the $J=7$--$6$ transition of C$^{34}$S was also 
detected.
In all observed cases the ratio of the 
$J=7$--$6$ versus $J=5$--$4$ lines implies a warm,
($T > 50\,$K) and dense ($n > 10^{6}$) environment (Table \ref{t_cs}). 
The C$^{34}$S-derived temperature is higher than that derived from
H$_2$CO and more likely refers to the inner warm and dense region
rather than the outer envelope.
If more transitions are observed, a detailed density structure for each
of the sources could be determined.

\subsubsection{SO}

The only observed SO transition was unfortunately blended with a methanol
line and thus difficult to accurately measure. Despite this
complication, three sources, spanning a range of energetics and conditions,
have unambiguous detections for SO (MMS6, NGC2071, and PDR1).
Assuming the formaldehyde derived properties for the physical conditions
of the gas, RADEX was used to determine the required column density of
SO for each of the sources, The RADEX calculations for SO include 
energy levels up to $580\,$K.
The results are presented in Table \ref{t_cs}, along with the abundance
of SO in each source and the SO to CS abundance ratio.

Observations of more SO lines and other sulphur-bearing species, such
as SO$_2$ and H$_2$S, are needed to study possible differences in the
sulphur-chemistry in these sources.

\section{Discussion and Conclusion}
\label{sect:disc}

This molecular line study was undertaken to determine if morphological 
clues and qualitative indicators were observable across a range of 
environmental conditions, from pre-stellar and young protostellar envelopes 
$L_{\rm bol} \sim 1-100 L_\odot$ through infrared bright energetic sources 
$L_{\rm bol} \sim 400 L_\odot$, a PDR knot and a shock front.
Several trends are apparent, especially in the derived abundances of
many molecular species. As well, there are a number of spectral hints 
that protostellar sources reside within the sub-millimeter clumps,
excluding the PDR and the shock knot.

\subsection{Abundances}

Despite the inherent danger in assuming a single temperature and density
for the environments of each source, general trends in abundance
are observed. It
is worth noting that detailed modeling (van der Tak \etal\ 2000a, 
Sch\"oier \etal\ 2002, Doty \etal\ 2002, J\o rgensen \etal\ 2002) provides 
a much more accurate determination of relative abundances, especially for 
molecules which are excited only in parts of the envelope. 
Also, the presence of abundance gradients or `jumps' can be
established for some molecules. Such models require a determination of
the density and temperature profiles of the sources from the dust
continuum emission, which has not yet been completed for this study. 
Sch\"oier \etal\ (2002) show that the inferred abundances
using a constant temperature and density do not differ by more than a
factor of a few from the detailed analysis as long as the adopted
conditions are appropriate for the particular molecule or line, which
should be the case for this work.  However, detailed modeling
requires significant input as to the density and temperature profile of the
source, as well as relying on additional assumptions such as the dust 
emissivity profile. {Additionally, the Orion sources are bathed in a strong
external radiation field (Li \etal\ 2003) requiring careful consideration
of the exterior conditions of each source envelope where the dust and gas 
temperatures reach $T \sim 30\,K$. This study is concerned primarily with the 
constraints on physical conditions provided by the variations in the molecular 
tracers without resorting to detailed modeling of individual sources.}

Some strong trends are observed across the source list. The peak brightness
and integrated line strength of both CO and $^{13}$CO 3--2 lines follow the
energetics and warm dust temperatures of the envelopes. Despite the observed
depletion of CO in the protostellar source envelopes (typically the abundance 
is a factor of 10 lower than the mean molecular cloud value), in all cases 
the objects are visible in each of the observed CO lines.

{The formaldehyde derived temperatures, T(H$_2$CO), do not match the envelope
dust temperatures, measured in the sub-millimeter.
However, for the sub-millimeter sources MMS6, MMS7, and MMS9 T(H$_2$CO)
is consistent with the gas temperature in the outer
envelope as seen in the $^{13}$CO line.  For the infrared sources, FIR4 and NGC2071,
T(H$_2$CO) requires internal heating through a significant
fraction of each envelope.}
Detailed modeling of other
sources provides evidence that the formaldehyde abundance increases in the 
warm, dense interiors of protostellar envelopes, and that grain mantle 
evaporation may be important in producing the enhancement.
However, the relatively low lying lines observed in this study are more
accurate tracers of the extended envelope conditions, where most of the mass 
resides (J\o rgensen \etal\ in prep.).  {While more uncertain,
the methanol derived temperatures follow the same trend as the formaldehyde
temperatures.} For the remainder of the discussion 
we adopt the formaldehyde conditions as representative of the bulk conditions 
within each source. 

{Despite uncertainties of order unity,} the abundance of both 
formaldehyde and methanol correlate well against the 
source energetics (Fig.\ \ref{f_abund}).  The changing abundance
may reflect jump conditions in formaldehyde and methanol abundance with dust 
temperature in the inner warm part.  
Ignoring the broad (outflow) methanol component in
NGC2071, the abundance of both formaldehyde and methanol may be plotted as a 
step function with cold sources, $T < 50\,$K, having a low abundance and 
warm sources, $T > 50\,$K, having about five times higher abundance.
Although less observational data from the study are available, the CS
abundance also appears to correlate with source energetics.

The abundance of HCN is also sensitive to source conditions; however, 
this is mostly determined by extreme cases MMS6 and PDR1.  In contrast
to the other observed species, little variation of abundance was found
for HCN among the energetic sources, FIR4 and NGC2071, and the weak 
sub-millimeter source MMS9.
Considering the three dominant
CN-bearing species (HCN, HNC, and CN), the total abundance of these three
molecules is quite constant across a wide range of conditions, varying
by less than thirty percent between MMS9, FIR4, and NGC2071 and only varying
by a factor of two when the PDR region is included. Only the cold, dense
sub-millimeter source MMS6 appears to have a severely depleted abundance of
CN-bearing species. {The enhanced external radiation field in Orion may be
responsible for this apparent equilibrium.}

{The general observed trend of increased abundance with 
source energetics is consistent with a scenario 
in which freeze-out of molecules occurs in the cooler intermediate zone 
within the envelope and evaporation of ices occurs in the warmer interior 
and possibly exterior regions.} Similar trends have been observed for
samples of low-mass (J{\o}rgensen \etal\ 2002, 2003) and high-mass
(van der Tak \etal\ 2000a,b; Boonman \etal\ 2003) 
sources without external heating, 
although not for all molecules.  The only other intermediate-mass source 
that has been studied in some
detail chemically is AFGL~490 (Schreyer \etal\ 2002) which has a
luminosity of $10^3$ L$_{\odot}$. This source has a large envelope,
with abundances comparable to those of the warmer sources observed
here. Strong solid-state features of various species indicative of
freeze-out are also detected.

None of the observed sources shows the characteristic crowded line
spectra of a `hot core', such as found for high-mass protostars like
G34.3+0.15 (Macdonald \etal\ 1996), G327.3--0.6 (Gibb \etal\ 2000)
or W3(H$_2$O) (Helmich \& van Dishoeck 1997).  Deeper integrations
are needed to determine whether complex organic molecules like
CH$_3$OCH$_3$ are present in the warmer sources in our sample,
e.g. FIR4 or NGC 2071. These molecules are expected to be produced by
gas-phase chemistry between evaporated ices, but have so far only
been observed in high-mass sources. The different dynamical time
scales of the hot core gas in low- versus high-mass objects compared
with the chemical timescales of $\sim 10^4$ yr may prevent formation
of such second generation species.

The PDR1 position differs chemically from the other sources by having
the highest gas-phase abundances overall and the largest [HCN]/[HNC]
ratio.  PDRs also generally have high abundances of radicals such as
CN and C$_2$H and ions like CO$^+$ (Jansen \etal\ 1995). Indeed, a
study of a set of more evolved intermediate-mass sources by Fuente et
al.\ (1998) has identified CN as a particularly good diagnostic of the
ultraviolet radiation. In our sources, this trend is not so evident,
likely because all sources are located in the Orion region which is
bathed in intense radiation from nearby O and B stars.

Few molecules are observed in the SK1-OMC3 shock and only CO shows
truly broad line wings at that position.  Sulphur-containing species
like H$_2$S, SO and SO$_2$ are predicted to be abundant in shock
models (e.g., Leen \& Graff 1988) and are seen to be enhanced in the
Orion-KL plateau gas with broad line wings, but are not prominent
here; however, the source is quite weak and the lines may be below
our detection limits. In NGC 2071,
molecules like CH$_3$OH are present in the outflow but this results
from grain evaporation in the shock rather than high-temperature
chemistry. Chernin, Mason, \& Fuller (1994) have observed broad SO emission in
NGC 2071 with some abundance enhancement.  Deeper searches for shock
diagnostics and accurate determinations of abundances at shocked and
non-shocked positions in these and other sources are needed.

\subsection{Protostellar Source Diagnostics}

A large number of sub-millimeter continuum surveys have now been completed in 
star-forming molecular clouds, and hundreds of new sub-millimeter bright envelopes have
been enumerated. However, an outstanding question is which fraction of these
objects surround protostellar sources and which are pre-stellar. Within the 
present study there were no clear pre-stellar objects although MMS6 appear
somewhat ambiguously defined as a Class 0 source. Despite this lack of an obvious
pre-stellar candidate, it is worth considering if any of the observed molecular
signatures {\it require} an embedded heating source.

For many of the sub-millimeter objects found within surveys, $^{13}$CO
observations have provided no indication of a coincident CO peak
(Mitchell \etal\ 2001). Thus, the clear measurement of CO isotopes in this sample, 
despite the depletion, may provide a clue to which objects contain embedded sources,
perhaps by warming and evaporating CO back into the environment from which it
had frozen out during an earlier epoch. {Alternatively, the presence
of a strong external radiation field may bias the results in Orion.} 
The broad CO line wings provide
clear evidence of enhanced kinematics within these condensations and, along
with the ubiquitous outflows in star-forming regions, act as a sign post
for embedded sources.

The appearance of molecular lines with relatively high excitation
temperatures, and preferably also with high critical densities such that the
warm region might be inferred to be deeper within the envelope, should provide
a signature to Class 0 and later sources. Thus, the methanol lines are likely 
the strongest indicator of a warm, dense region within the envelope. For all
sources observed in this study, except the known outflow source MMS9, the
required internal temperature deduced from the methanol lines is $> 50\,$K,
a temperature unattainable without an energetic internal source. 
The CS 7--6 line also provides evidence of a warm, dense interior region
in all sources except MMS9. It is interesting to note that all the protostellar
sources considered in this paper require high densities in their interior. 
Standard models for star formation expect a power-law density distribution 
toward the clump center during the collapse phase (Shu 1977; Henriksen, Andr\'e,
\& Bontemps 1997) The high densities measured here may be a result of
the stage of evolution of the individual clumps. 

{Formaldehyde observations also provide important evidence for a large
warm envelope. While the low-lying lines observed in this study are
predominantly excited in the massive outer envelope, the derived
temperature places constraints on the required heating source.
In particular, envelope temperatures above $20\,$K are
difficult to reconcile without either a warming source inside or an external 
heating source.  All the sources observed in this study have two components,
narrow and broad, associated with the formaldehyde measurements. However, this 
is not always the case for sub-millimeter bright regions (Mitchell \etal\ 2001; 
Tothill, private communication).
It is possible that the broad component of the formaldehyde, often implying
a much denser zone, is tracing an inner region within the envelope which
is undergoing collapse or has become much more turbulent.}

Not all heating is due to protostellar sources and it is therefore important
to distinguish external heating (such as in a PDR) from internal heating. 
Protostellar sources should have central condensations which show up both
in sub-millimeter dust continuum maps and maps of molecular lines with high 
critical densities (e.g., CS 7-6). Chemically, PDRs are best recognized by a 
high [HCN]/[HNC] ratio and by high abundances of radicals such as CN and 
C$_2$H and ions like CO$^+$ (see above discussion).

\subsection{Line Contamination}

One of the original motivations for this molecular line study was to determine 
the importance of line contamination within the broad SCUBA 850\,$\mu$m 
passband. While theoretical calculations (Appendix A) show that the influence
of CO contamination can become exceedingly large in particular situations,
it is also clear that typical conditions within molecular clouds are not
so extreme. The observational evidence presented (Table \ref{t_cont}) 
does show that the contamination, while typically less than 10\%, occasionally 
{\it dominates} the continuum flux.  However, this only occurs in regions with
warmer molecular gas temperatures {\it and} large velocity gradients
which allow for enhanced integrated line strengths in CO and its isotopes.
Such conditions occur most often at isolated shock fronts within the cloud, 
for example in the knots associated with protostellar jets.
Contamination may also arise from lines of other molecules,
especially around energetic sources. The best observed example for this
is Orion KL, for which a forest of lines, primarily from SO and SO$_2$, produce
between 28\% and 50\% line contamination at 850\,$\mu$m 
(Serabyn \& Weisstein 1995; Groesbeck 1994).
Consideration of Table \ref{t_cont} shows that in our sample the line 
contamination is {\it never} dominated by lines other than CO; however,
for the most energetic sources line emission from HCN, HNC, CN, and
methanol provides a significant fraction ($> 40\,$\%) of the total line
contamination.

\begin{acknowledgements}
We wish to thank Jes J\o rgensen, Floris van der Tak, John Bally and Heather Scott for 
helpful comments during the creation of this manuscript. The careful review by the
anonymous referee produced many improvements to the paper.
DJ wishes to thank the Sterrewacht Leiden for its kind hospitality
during the past three summers, at which time most of the data
analysis was conducted. DJ also acknowledges the support of an
NSERC grant and support from NOVA. This work was supported by the
Netherlands Organization for Scientific Research (NWO) through
grant 614-041-003 and a Spinoza grant.
\end{acknowledgements}

\vfill
\clearpage

\input 3536_t1.tex
\clearpage
\input 3536_t2.tex
\clearpage
\input 3536_t3.tex
\clearpage
\input 3536_t4.tex
\clearpage
\input 3536_t5.tex
\clearpage
\input 3536_t6.tex
\clearpage
\input 3536_t7.tex
\clearpage
\input 3536_t8.tex
\clearpage
\input 3536_t9.tex
\clearpage
\input 3536_cap.tex
\clearpage

\appendix

\section{Contamination of sub-millimeter continuum by molecular lines}

Comparison of 850\,$\mu$m continuum against CO 3--2 measurements in the spiral 
arms of external disk galaxies has revealed that as much as 50\% 
of the emission within the SCUBA bandpass is contamination due to this 
single line (Tilanus, van der Werf, \& Israel 2000). 
This result is not surprising since the energy radiated 
in an optically thin  CO $j$-$i$ line by a column of gas with gas temperature 
$T_g$ and number density $n$ is given by
\begin{equation}
S_{ji} = \left( {h\, \nu_{ji} \over 4 \pi} \right)\, A_{ji}\, N(\rm{CO})\, 
X_j(T_g, n)
\ \ \ {\rm erg\ s}^{-1}{\rm cm}^{-2}{\rm sr}^{-1},
\end{equation}
where the column of CO is $N(\rm {CO}) = x({\rm CO})\,N(\rm{H}_2)$, and the
fractional population of the jth level is $X_j$.
For the CO 3--2 line, $\nu_{32} = 3.458 \times 10^{11}$ s$^{-1}$ and
$A_{32} = 2.48 \times 10^{-6}$ s$^{-1}$. 
Alternatively, for optically thin dust the
energy radiated across a passband centered at $\nu_d$ and with an
effective width $\Delta \nu$ is
\begin{equation}
S_d = B_{\nu_d}(T_d)\,N_{H_2}\,\mu\,m_{H}\,\kappa_{\nu_d}
\,\Delta \nu 
\ \ \ {\rm erg\ s}^{-1}{\rm cm}^{-2}{\rm sr}^{-1},
\end{equation}
where $\kappa_{\nu_d}$ is the emissivity of the dust at $\nu_d$ and
$\mu$ is the mean molecular weight of the gas.
Since the dust radiation in molecular clouds is usually near the 
Rayleigh-Jeans limit, one can replace the Planck function 
$B_{\nu}(T_d)$ with $2\,k\,T_d f_{\nu}(T_d)/ \lambda_{\nu}^2$,
where $f_{\nu}(T_d)$ accounts for the difference between the Planck law and
the Rayleigh-Jeans limit at frequency $\nu$.

Thus, in the optically thin limit the ratio of the line to continuum is
independent of the column of hydrogen, depending only on the temperature
and density within the region, the abundance of CO in the gas phase, and the
dust emissivity.
\begin{equation}
R = { 1 \over 8 \pi}    
\left({h \nu_{ji} \over  k T_d}\right)\,
\left({A_{ji} \over \Delta \nu}\right)\,
\left({\lambda_{\nu_d}^2\over \mu\,m_{H}\,\kappa_{\nu_d}}\right)\,
f_{\nu_d}(T_d)\,x({\rm CO})\,X_3(T_g, n).
\end{equation}
Computing the ratio of the CO 3--2 line against the JCMT 850\,$\mu$m continuum
band, where $\Delta \nu = 3\times 10^{10}\,$s$^{-1}$, yields
\begin{equation}
R_{850} = 28\,  
\left({T_d \over 20\,{\rm K}}\right)^{-1}\, 
\left({ \kappa_{850} \over 0.02 {\rm cm}^2/{\rm g}}\right)^{-1}\,
\left({ x({\rm CO}) \over 1 \times 10^{-4}}\right)\,
f_{850}(T_d)\,X_3(T_g, n).
\end{equation}
The ratio depends strongly on the population of the upper CO energy
level $X_3$ which is a function of both the gas temperature and the
density. Tables \ref{t_frac} and \ref{t_ratio} show $X_3$ and $R_{850}$ 
for a range of internal gas conditions, assuming that the gas 
dust temperatures are equal. The values shown in Table \ref{t_frac} 
{\it only} apply in the optically thin limit for the CO 3--2 transition
($N({\rm H}_2) < 10^{21}\,$cm$^{-3}$).  At higher columns, 
as are expected deep within molecular clouds, the continuum emission 
continues to increase linearly while the line emission is severely damped
and the contamination is expected to be less influential.
For extragalactic studies, where the spatial resolution does not
permit the separation of thick and thin portions of the cloud, it is
not surprising that the CO emission produces a significant contamination.

The degree to which the CO line contamination affects sub-millimeter
continuum measurements at high optical depth within molecular clouds, as
investigated in this paper, depends sensitively on the width of the 
CO 3--2 line and the strength of the extended line wings, and should be 
measured independently where possible. However, as long as the line is
not significantly broadened, the CO contamination should be minor. Freeze-out
of the CO onto dust grains decreases the importance of CO contamination
for the densest and coldest regions.
\eject

\input 3536_tA1.tex
\input 3536_tA2.tex

\clearpage

\end{document}

%% file: 3536_t1.tex
\begin{table}
\caption[]{Source List and Sub-millimeter Diagnostics}
$$
\begin{array}{lcccccc}
\hline
\noalign{\smallskip}
{\rm Name}&{\rm R.A.}^{[\mathrm{a}]}&{\rm Dec.^{[\mathrm{a}]}}&
T_d&M_{\rm env}&L&{\rm Object}\\
\noalign{\smallskip}
&{\rm (J2000)}&{\rm (J2000)}&(K)&(M_\odot)&(L_\odot)&\\
\noalign{\smallskip}
\hline
\noalign{\smallskip}
{\rm MMS6-OMC3}&{\rm 5:35:23.5}&{\rm -05:01:32}& 	 15&36&1.2-60&{\rm Class 0}  \\
{\rm MMS9-OMC3}&{\rm 5:35:26.2}&{\rm -05:05:46}&	 20&10&0.6-90&{\rm Class 0}  \\
{\rm MMS7-OMC3}&{\rm 5:35:26.4}&{\rm -05:03:59}&26-50& 8&    76&{\rm Class 0}  \\
{\rm FIR4-OMC2}&{\rm 5:35:26.7}&{\rm -05:09:59}&15-40&34&   400&{\rm Class 0/I}\\
{\rm NGC2071IR}&{\rm 5:47:04.4}&{\rm +00:21:49}&20-50&30&   500&{\rm Class 0/I}\\
\\
{\rm PDR1-OMC1}&{\rm 5:35:25.3}&{\rm -05:24:34}&	 60&&&	{\rm PDR}\\
{\rm SK1-OMC3} &{\rm 5:35:17.0}&{\rm -05:06:03}&	 10&&&	{\rm Shock}\\
\noalign{\smallskip}
\hline
\end{array}
\label{t_prop}
$$
\begin{list}{}{}
\item[$^{\mathrm{a}}$] Positions are taken from the sub-millimter maps of the 
ISF (Johnstone \& Bally 1999) and Orion B North (Johnstone \etal\ 2001).
\end{list}
\end{table}

%% file: 3536_t2.tex
\begin{table}
\caption[]{Calendar of Observations and Conditions}
$$
\begin{array}{lcccccc}
\hline
\noalign{\smallskip}
{\rm Date}&
{\rm Receiver}&
{\nu_{\rm upper}}&
{\nu_{\rm lower}}&
{\Delta V}&
{T_{\rm sys}}&
{\rm Lines}\\
\noalign{\smallskip}
{}&
{}&
{\rm (GHz)}&
{\rm (GHz)}&
{\rm (km/s)}&
{\rm (K)}&
{}\\
\noalign{\smallskip}
\hline
\noalign{\smallskip}
{\rm 1999\ January  }&{\rm RxA3}& 218.35& 226.48& 0.2& 400-700&{\rm H}_2{\rm CO}, {\rm CH}_3{\rm OH}\\
{\rm 1999\ September}&{\rm RxB3}& 330.53& 338.53& 0.5& 1200-1700&^{13}{\rm CO}, {\rm CH}_3{\rm OH}\\
{\rm 1999\ September}&{\rm RxB3}& 338.53& 346.53& 0.5& 1200-1700&{\rm CH}_3{\rm OH}, {\rm SO}\\
{\rm 1999\ September}&{\rm RxB3}& 354.40& 362.40& 0.5& 1200-1700&{\rm HCN, HNC}\\
{\rm 2000\ December} &{\rm RxA3}& 233.00& 241.00& 0.4& 350-400&{\rm C}^{34}{\rm S}\\ 
{\rm 2000\ December} &{\rm RxA3}& 233.75& 241.75& 0.4& 350-400&{\rm CH}_3{\rm OH}\\
{\rm 2000\ December} &{\rm RxA3}& 236.94& 244.94& 0.4& 375-385&{\rm CS}\\ 
{\rm 2000\ December} &{\rm RxB3}& 337.25& 345.25& 0.5& 530-550, 710-740&{\rm C}^{17}{\rm O}, {\rm C}^{34}{\rm S}\\
{\rm 2000\ December} &{\rm RxB3}& 340.45& 348.45& 0.5& 700-800&{\rm CN, HN}^{13}{\rm C}\\
{\rm 2000\ December} &{\rm RxB3}& 345.67& 353.67& 0.5& 490-510&{\rm CO}\\
{\rm 2000\ December} &{\rm RxB3}& 354.60& 362.60& 0.5& 590-620& {\rm HCN, HNC, H}_2{\rm CO}\\
\noalign{\smallskip}
\hline
\end{array}
$$
\label{t_cal}
\end{table}

%% file: 3536_t3.tex
\begin{table}
\caption{Observed Spectra and Gaussian Fit Parameters}
\begin{tabular}{llrrrrr}
\hline
\noalign{\smallskip}
{Source}&
{Species}&
{$\nu$}&
{J}&
{$W$}&
{$\Delta V$}&
{$T_{\rm MB}$}\\
\noalign{\smallskip}
{}&
{}&
{(GHz)}&
{}&
{(K$\,$km\,s$^{-1}$)}&
{(km\,s$^{-1}$)}&
{(K)}\\
\hline
\noalign{\smallskip}
MMS6-OMC3&CO       &345.796&3--2			   &93.00& -- & -- \\
&$^{13}$CO&330.588&3--2                    &4.22&0.57&7.00\\
&         &       &                        &40.70&2.24&17.1\\
&C$^{17}$O&337.061&3--2                     &4.78&1.08&4.15\\
\\
&H$_2$CO  &218.222&3$_{03}$--2$_{02}$       &2.81&0.88&3.00\\
&	  &       &                         &2.52&2.40&0.99\\
&H$_2$CO  &218.475&3$_{22}$--2$_{21}$       &0.53&1.73&0.29\\
&H$_2$CO  &362.736&5$_{05}$--4$_{04}$       &1.80&0.91&1.85\\
&         &       &                         &1.76&2.92&0.57\\
\\
&CH$_3$OH &218.440&4$_2$E-3$_1$E            &0.44&1.41&0.29\\
&CH$_3$OH &241.700&5$_0$E-4$_0$E            &0.18&0.82&0.21\\
&CH$_3$OH &241.767&5$_{-1}$E-4$_{-1}$E      &0.95&1.63&0.55\\
&CH$_3$OH &241.791&5$_0$A$^+$-4$_0$A$^+$    &1.27&1.65&0.72\\
&CH$_3$OH &241.904&5$_{\pm2}$E-4$_{\pm2}$E  &0.34&2.80&0.11\\
&CH$_3$OH &338.345&7$_{-1}$E-6$_{-1}$E      &0.84&1.56&0.50\\
&CH$_3$OH &338.404&7$_6$E-6$_6$E            &0.34&0.99&0.32\\
&         &338.409&7$_0$A$^+$-6$_0$A$^+$    &0.96&3.45&0.26\\
&CH$_3$OH &338.583&7$_3$E-6$_3$E            &0.51&4.81&0.10\\
&CH$_3$OH &338.615&7$_1$E-6$_1$E            &0.46&2.05&0.21\\
&CH$_3$OH &338.722&7$_{\pm2}$E-6$_{\pm2}$E  &0.45&1.69&0.25\\
\\
&CN       &340.248&3-2                      &2.95&1.41&1.97\\
&HNC      &362.630&4-3                      &4.56&1.06&4.04\\
&HCN      &354.505&4-3                      &9.65&3.81&2.38\\
&H$^{13}$CN&345.340&4-3                     &0.49&2.05&0.23\\
\\
&CS       &244.936&5-4                      &2.31&0.93&2.34\\
&         &       &                         &2.23&2.85&0.74\\
&C$^{34}$S&241.016&5-4                      &0.65&1.58&0.39\\
&C$^{34}$S&337.396&7-6                      &0.54&1.19&0.43\\
&SO       &346.528&8,9-7,8                  &1.28&1.23&0.98\\
\\
\hline
\noalign{\smallskip}
\end{tabular}
\addtocounter{table}{-1}
\label{t_spec}
\end{table}
\clearpage
\begin{table}
\caption{Observed Spectra and Gaussian fit parameters (cont'd)}
\begin{tabular}{lcrrrrr}
\hline
\noalign{\smallskip}
{Source}&
{Species}&
{$\nu$}&
{J}&
{$W$}&
{$\Delta V$}&
{$T_{\rm MB}$}\\
\noalign{\smallskip}
{}&
{}&
{(GHz)}&
{}&
{(K$\,$km\,s$^{-1}$)}&
{(km\,s$^{-1}$)}&
{(K)}\\
\hline
\noalign{\smallskip}
MMS7-OMC3&H$_2$CO  &218.222&3$_{03}$--2$_{02}$       &2.03&1.38&1.39\\
&H$_2$CO  &218.475&3$_{22}$--2$_{21}$       &0.10&0.55&0.18\\
\\
MMS9-OMC3&CO       &345.796&3--2                   &225.00& -- & -- \\
&$^{13}$CO&330.588&3--2                    &59.10&2.40&23.2\\
&C$^{17}$O&337.061&3--2                     &2.88&1.29&2.10\\
\\
&H$_2$CO  &218.222&3$_{03}$--2$_{02}$       &1.42&0.67&1.99\\
&         &       &                         &1.65&2.11&0.73\\
&H$_2$CO  &218.475&3$_{22}$--2$_{21}$       &0.28&0.74&0.36\\
&H$_2$CO  &362.736&5$_{05}$--4$_{04}$       &0.94&1.12&0.78\\
\\
&CH$_3$OH &218.440&4$_2$E-3$_1$E            &0.23&0.98&0.22\\
&CH$_3$OH &241.700&5$_0$E-4$_0$E            &0.14&0.82&0.16\\
&CH$_3$OH &241.767&5$_{-1}$E-4$_{-1}$E      &0.51&1.21&0.40\\
&CH$_3$OH &241.791&5$_0$A$^+$-4$_0$A$^+$    &0.66&0.86&0.72\\
&CH$_3$OH &338.345&7$_{-1}$E-6$_{-1}$E      &0.30&0.55&0.50\\
\\
&CN       &340.248&3-2                      &4.65&1.78&2.46\\
&HNC      &362.630&4-3                      &2.97&0.98&2.85\\
&HCN      &354.505&4-3                      &1.05&0.82&1.21\\
&         &       &                         &3.57&3.40&0.98\\
&H$^{13}$CN&345.340&4-3                     &0.41&5.23&0.07\\
\\
&CS       &244.936&5-4                      &2.73&1.50&1.71\\
\\
\hline
\noalign{\smallskip}
\end{tabular}
\label{t_spec_2a}
\addtocounter{table}{-1}
\end{table}
\clearpage
\begin{table}
\caption{Observed Spectra and Gaussian fit parameters (cont'd)}
\begin{tabular}{lcrrrrr}
\hline
\noalign{\smallskip}
{Source}&
{Species}&
{$\nu$}&
{J}&
{$W$}&
{$\Delta V$}&
{$T_{\rm MB}$}\\
\noalign{\smallskip}
{}&
{}&
{(GHz)}&
{}&
{(K$\,$km\,s$^{-1}$)}&
{(km\,s$^{-1}$)}&
{(K)}\\
\hline
\noalign{\smallskip}
FIR4-OMC2&CO       &345.796&3--2			  &442.00& -- & -- \\ 
&$^{13}$CO&330.588&3--2                    &76.20&2.57&27.8\\
&         &       &                        &24.70&12.5&1.85\\
&C$^{17}$O&337.061&3--2                     &1.27&0.90&1.32\\
&         &       &                         &2.70&1.65&1.54\\
\\
&H$_2$CO  &218.222&3$_{03}$--2$_{02}$       &9.82&1.49&6.21\\
&	  &       &                        &15.70&6.28&2.35\\
&H$_2$CO  &218.475&3$_{22}$--2$_{21}$       &2.96&1.50&1.85\\
&	  &       &                         &4.49&5.81&0.73\\
&H$_2$CO  &362.736&5$_{05}$--4$_{04}$       &5.24&1.20&4.09\\
&         &       &                         &29.3&6.02&4.50\\
\\
&CH$_3$OH &218.440&4$_2$E-3$_1$E            &3.43&0.98&3.29\\
&         &       &                         &7.49&5.10&1.38\\
&CH$_3$OH &241.700&5$_0$E-4$_0$E            &2.39&1.36&1.65\\
&         &       &                         &5.36&5.33&0.95\\
&CH$_3$OH &241.767&5$_{-1}$E-4$_{-1}$E      &4.73&1.29&3.44\\
&         &       &                         &9.33&5.50&1.60\\
&CH$_3$OH &241.791&5$_0$A$^+$-4$_0$A$^+$    &5.93&1.36&4.11\\
&         &       &                        &10.10&5.25&1.81\\
&CH$_3$OH &241.904&5$_{\pm2}$E-4$_{\pm2}$E  &2.35&1.54&1.43\\
&         &       &                         &6.65&5.68&1.10\\
&CH$_3$OH &338.345&7$_{-1}$E-6$_{-1}$E      &5.86&1.34&4.12\\
&         &       &                        &21.70&5.28&3.85\\
&CH$_3$OH &338.404&7$_6$E-6$_6$E            &5.31&1.21&4.14\\
&         &       &                        &25.20&5.30&4.46\\
&CH$_3$OH &338.513&7$_4$A$^\pm$-6$_4$A$^\pm$&1.58&0.57&2.59\\
&         &       &                        &13.30&7.37&1.69\\
&CH$_3$OH &338.542&7$_3$A$^\pm$-6$_4$A$^\pm$&5.37&2.29&2.20\\
&         &       &                        &11.30&3.54&3.00\\
&CH$_3$OH &338.583&7$_3$E-6$_3$E            &1.44&3.03&0.45\\
&         &       &                         &1.96&5.70&0.32\\
&CH$_3$OH &338.615&7$_1$E-6$_1$E            &2.24&1.26&1.67\\
&         &       &                        &10.20&5.68&1.69\\
&CH$_3$OH &338.640&7$_2$A$^+$-6$_20$A$^+$   &9.44&4.19&2.12\\
&CH$_3$OH &338.722&7$_{\pm2}$E-6$_{\pm2}$E &11.60&2.70&4.03\\
&         &       &                        &29.20&7.23&3.79\\
\\
&CN       &340.248&3-2                      &7.48&1.88&3.73\\
&         &       &                        &15.50&8.68&1.67\\
&HNC      &362.630&4-3                      &9.63&1.62&5.60\\
&         &       &                        &11.50&6.83&1.58\\
&HCN      &354.505&4-3                    &163.00&10.2&15.0\\
&H$^{13}$CN&345.340&4-3                     &8.66&8.43&0.97\\
\\
&C$^{34}$S&241.016&5-4                      &0.72&1.40&0.48\\
&         &       &                         &1.31&5.18&0.24\\
&C$^{34}$S&337.396&7-6                      &2.64&7.39&0.34\\
&SO       &346.528&8,9-7,8                  &$<$9.00&6.00&0.97\\
&         &       &                         &6.70&6.00&1.10\\
\\
\hline
\noalign{\smallskip}
\end{tabular}
\addtocounter{table}{-1}
\label{t_spec_3}
\end{table}
\clearpage
\begin{table}
\caption{Observed Spectra and Gaussian fit parameters (cont'd)}
\begin{tabular}{lcrrrrr}
\hline
\noalign{\smallskip}
{Source}&
{Species}&
{$\nu$}&
{J}&
{$W$}&
{$\Delta V$}&
{$T_{\rm MB}$}\\
\noalign{\smallskip}
{}&
{}&
{(GHz)}&
{}&
{(K$\,$km\,s$^{-1}$)}&
{(km\,s$^{-1}$)}&
{(K)}\\
\hline
\noalign{\smallskip}
NGC2071IR&CO       &345.796&3--2			  &770.00& -- & -- \\
&C$^{17}$O&337.061&3--2                     &2.96&1.73&1.60\\
&         &       &                         &4.25&7.31&0.55\\
\\
&H$_2$CO  &218.222&3$_{03}$--2$_{02}$      &12.00&2.21&5.10\\
&	  &       &                        &14.00&9.12&1.45\\
&H$_2$CO  &218.475&3$_{22}$--2$_{21}$       &3.75&2.95&1.19\\
&	  &       &                         &3.67&16.3&0.21\\
&H$_2$CO  &362.736&5$_{05}$--4$_{04}$      &10.90&2.07&4.93\\
&         &       &                        &10.70&7.50&1.34\\
\\
&CH$_3$OH &218.440&4$_2$E-3$_1$E            &3.12&2.30&1.28\\
&CH$_3$OH &241.700&5$_0$E-4$_0$E            &2.39&1.36&1.65\\
&         &       &                         &5.36&5.33&0.94\\
&CH$_3$OH &241.767&5$_{-1}$E-4$_{-1}$E      &6.51&2.38&2.56\\
&         &       &                         &6.18&12.1&0.48\\
&CH$_3$OH &241.791&5$_0$A$^+$-4$_0$A$^+$    &4.93&1.80&2.58\\
&         &       &                         &8.12&7.25&1.05\\
&CH$_3$OH &241.879&5$_{1}$E-4$_{1}$E        &1.45&1.88&0.73\\
&CH$_3$OH &241.888&5$_2$A$^+$-4$_2$A$^+$    &0.74&2.24&0.31\\
&CH$_3$OH &241.904&5$_{\pm2}$E-4$_{\pm2}$E  &1.09&1.52&0.67\\
&         &       &                         &2.98&4.66&0.60\\
\\
&CN       &340.248&3-2                     &14.50&2.61&5.20\\
&         &       &                        &14.50&8.63&1.58\\
&HNC      &362.630&4-3                     &12.30&1.92&6.00\\
&         &       &                        &10.70&6.44&1.56\\
&HN$^{13}$C&345.340&4-3                     &1.30&3.89&0.31\\
&HCN      &354.505&4-3                     &52.10&3.95&12.4\\
&         &       &                        &31.20&14.3&2.05\\
&H$^{13}$CN&345.340&4-3                     &5.51&10.1&0.51\\
\\
&CS       &244.936&5-4                     &10.30&1.87&5.17\\
&         &       &                        &25.50&6.29&3.81\\
&C$^{34}$S&241.016&5-4                      &0.59&0.85&0.66\\
&         &       &                         &3.86&3.45&1.05\\
&C$^{34}$S&337.396&7-6                      &0.53&3.70&0.13\\
&	  &       &                         &3.57&3.74&0.89\\
\\
\hline
\noalign{\smallskip}
\end{tabular}
\addtocounter{table}{-1}
\label{t_spec_4}
\end{table}
\clearpage
\begin{table}
\caption{Observed Spectra and Gaussian fit parameters (cont'd)}
\begin{tabular}{lcrrrrr}
\hline
\noalign{\smallskip}
{Source}&
{Species}&
{$\nu$}&
{J}&
{$W$}&
{$\Delta V$}&
{$T_{\rm MB}$}\\
\noalign{\smallskip}
{}&
{}&
{(GHz)}&
{}&
{(K$\,$km\,s$^{-1}$)}&
{(km\,s$^{-1}$)}&
{(K)}\\
\hline
\noalign{\smallskip}
PDR1-OMC1&CO       &345.796&3--2			  &535.00& -- & -- \\
&$^{13}$CO&330.588&3--2                   &129.00&1.39&87.0\\
&         &       &                        &90.10&1.28&66.0\\
&C$^{17}$O&337.061&3--2                    &10.20&1.38&6.94\\
\\
&H$_2$CO  &218.222&3$_{03}$--2$_{02}$       &0.38&0.49&0.72\\
&	  &       &                         &5.11&1.88&2.55\\
&H$_2$CO  &218.475&3$_{22}$--2$_{21}$       &0.23&0.49&0.43\\
&	  &       &                         &1.76&1.32&1.25\\
&H$_2$CO  &362.736&5$_{05}$--4$_{04}$       &3.80&1.56&2.29\\
\\
&CH$_3$OH &241.700&5$_0$E-4$_0$E            &0.33&1.83&0.17\\
&CH$_3$OH &241.767&5$_{-1}$E-4$_{-1}$E      &0.89&1.87&0.45\\
&CH$_3$OH &241.791&5$_0$A$^+$-4$_0$A$^+$    &1.20&1.83&0.61\\
&CH$_3$OH &241.904&5$_{\pm2}$E-4$_{\pm2}$E  &0.35&1.71&0.19\\
&CH$_3$OH &338.345&7$_{-1}$E-6$_{-1}$E      &1.14&1.43&0.75\\
&CH$_3$OH &338.404&7$_6$E-6$_6$E            &1.85&1.86&0.94\\
&CH$_3$OH &338.722&7$_{\pm2}$E-6$_{\pm2}$E  &0.74&2.03&0.34\\
\\
&CN       &340.248&3-2                     &23.80&2.18&10.3\\
&HNC      &362.630&4-3                      &3.83&1.57&2.29\\
&HCN      &354.505&4-3                     &41.70&2.39&16.4\\
&H$^{13}$CN&345.340&4-3                     &2.11&1.47&1.35\\
\\
&C$^{34}$S&241.016&5-4                      &7.32&1.65&4.16\\
&C$^{34}$S&337.396&7-6                      &5.74&1.28&4.21\\
&SO       &346.528&8,9-7,8                  &4.82&1.64&2.77\\
\\
SK1-OMC3&CO       &345.796&3--2			  &404.00& -- & -- \\
&$^{13}$CO&330.588&3--2                     &5.33&1.12&4.49\\
&         &       &                        &69.50&2.49&26.2\\
&C$^{17}$O&337.061&3--2                     &3.31&1.60&1.94\\
\\
&H$_2$CO  &218.222&3$_{03}$--2$_{02}$       &0.57&1.45&0.37\\
&H$_2$CO  &218.475&3$_{22}$--2$_{21}$       &$<$0.20&0.49&0.37\\
\\
&CN       &340.248&3-2                      &0.94&3.49&0.25\\
&HNC      &362.630&4-3                      &0.23&1.25&0.17\\
&HCN      &354.505&4-3                      &0.43&1.24&0.32\\
&         &       &                         &1.74&14.6&0.11\\
\hline
\noalign{\smallskip}
\end{tabular}
\label{t_spec_6}
\end{table}

%% file: 3536_t4.tex
\begin{table}
\caption{Physical Conditions Discerned from Carbon-Monoxide}
\begin{tabular}{lcccccccccc}
\hline
\noalign{\smallskip}
{}&
{CO}&
{}&
{$^{13}$CO}&
{}&
{C$^{17}$O}&
{}&
{}&
{}&
{}\\
\noalign{\smallskip}
{Source}&
{$T_{\rm peak}$}&
{$T\Delta V$}&
{$T_{\rm peak}$}&
{$T\Delta V$}&
{$T_{\rm peak}$}&
{$T\Delta V$}&
{$N(\rm CO)$}&
{$\tau_{\rm mod}$}&
{$\tau_{\rm obs}/\tau_{\rm mod}$}\\
{}&
{(K)}&
{(K$\,$km\,s$^{-1}$)}&
{(K)}&
{(K$\,$km\,s$^{-1}$)}&
{(K)}&
{(K$\,$km\,s$^{-1}$)}&
{($10^{18}\,$cm$^{-2}$)}&
{$^{13}$CO}&
{}\\
\hline
\noalign{\smallskip}
MMS6&23.&93.&21.&45.&3.6&4.8&4.6&2.1&2.4\\
MMS9&28.&225&23.&59.&2.1&3.0&2.7&1.3&1.5\\
FIR4&42.&442&28.&101&2.7&4.1&3.4&1.8&1.3\\
NGC2071&67.&770&-- &-- &2.2&7.3&5.7&1.5&1.0\\
\\
PDR1&165&535&105&220&7.2&10.&11.&2.0&1.1\\
SK1&52.&404&26.&75.&1.9&3.3&2.8&1.5&0.9\\
\hline
\noalign{\smallskip}
\end{tabular}
\label{t_co}
\end{table}

%% file: 3536_t5.tex
\begin{table}
\caption{Measured Total Line Intensity within the 850$\,\mu$m Band and
Contribution to Continuum Flux}
\begin{tabular}{lccccc}
\hline
\noalign{\smallskip}
{Source}&
{CO Line Strength$^{[\mathrm{a}]}$}&
{Total Line Strength$^{[\mathrm{b}]}$}&
{Contamination}&
{$N(\rm{H}_2)$}&
{[CO]/[H$_2$]}\\
\noalign{\smallskip}
{}&
{(K$\,$km\,s$^{-1}$)}&
{(K$\,$km\,s$^{-1}$)}&
{\%}&
{($10^{22}$ cm$^{-2}$)}&
{($10^{-4}$)}\\
\hline
\noalign{\smallskip}
MMS6&143&170&1.5&104&0.04\\
MMS9&287&301&8.2& 14&0.19\\
MMS7&   &   &   & 22&\\
FIR4&547&934&8.0& 47&0.07\\
NGC2071&777&945$^{[\mathrm{c}]}$&8.3&62&0.09\\
\\
PDR1  &765&853&11.1& 10&1.10\\
SK1  &483&487 & 63.&  6&0.47\\
\hline
\noalign{\smallskip}
\end{tabular}
\label{t_cont}
\begin{list}{}{}
\item[$^{\mathrm{a}}$]
CO line strengths include contribution from CO, 
$^{13}$CO, and C$^{17}$O $J=3$--$2$ transitions.
\item[$^{\mathrm{b}}$]
Total line strength includes all measured lines within the
850$\mu$m passband.
\item[$^{\mathrm{c}}$]
The total line strength for NGC2071 does not include contributions
from methanol. Comparison with FIR4 suggests that an additional few 
percent of line contamination might be produced by these unobserved lines.
\end{list}
\end{table}

%% file: 3536_t6.tex
\begin{table}
\caption{Physical Conditions and Abundances Discerned from Formaldehyde}
\begin{tabular}{lcccccccc}
\hline
\noalign{\smallskip}
{Source}&
{C$^{[\mathrm{a}]}$}&
{$R_{33}^{[\mathrm{b}]}$}&
{$R_{53}^{[\mathrm{b}]}$}&
{$T$}&
{$n$}&
{$N$(H$_2$CO)}&
{[H$_2$CO]/[CO]}&
{[H$_2$CO]/[H$_2$]}\\
\noalign{\smallskip}
{}&
{}&
{}&
{}&
{(K)}&
{($10^6\,$cm$^{-3}$)}&
{($10^{13}\,$cm$^{-2}$)}&
{($10^{-5}$)}&
{($10^{-10}$)}\\
\hline
\noalign{\smallskip}
MMS6&n&$ >5$&0.6&$<50$&$1-3$&4.0&&\\
MMS6&b&$ >5$&0.6&$<50$&$1-3$&4.0&&\\
MMS6&t&     &   &     &    &8.0&1.7&0.8\\
\\
MMS9&t&  10&0.3&   30&   $1-3$&2.2&0.8&1.6\\
\\
MMS7&t&$ >7$& - &$ <40$&  - &2.0&&0.9\\
\\
FIR4&n&  3.3&0.7&   90& $0.5-2$&12.0&&\\
FIR4&b&  3.4&1.8&   80&  $2-10$&20.0&&\\
FIR4&t&     &   &     &    &32.0&9.4&6.5\\
\\
NGC2071&n&  3.2&0.9&  90&  $1-3$&15.0&&\\
NGC2071&b&  3.8&0.8&  70&  $1-3$&17.0&&\\
NGC2071&t&     &   &    &    &32.0&4.3&5.1\\
\\
PDR1  &n&  1.8& - &$>200$&-  &0.8&&\\
PDR1  &b&  2.8&0.7&  120&   $0.5-2$&6.0&&\\
PDR1  &t&     &   &     &    &6.8&0.6&6.8\\
\hline
\noalign{\smallskip}
\end{tabular}
\label{t_h2co}
\begin{list}{}{}
\item[$^{\mathrm{a}}$]
The letter designations refer to the velocity components with 
narrow (n), broad (b), and total (t).
\item[$^{\mathrm{b}}$]
$R_{33}$ is the ratio of the $3_{03}$--$2_{02}$ transition versus
the $3_{22}$--$2_{21}$ transition and 
$R_{53}$ is the ratio of the $5_{05}$--$4_{04}$ transition versus
the $3_{03}$--$2_{02}$ transition.
\end{list}
\end{table}

%% file: 3536_t7.tex
\begin{table}
\caption{Physical Conditions and Abundances Discerned from Methanol}
\begin{tabular}{lcccccccc}
\hline
\noalign{\smallskip}
{Source}&
{C$^{[\mathrm{a}]}$}&
{$T$}&
{$n$}&
{$N$(CH$_3$OH)}&
{[CH$_3$OH]/[CO]}&
{[CH$_3$OH]/[H$_2$]}\\
\noalign{\smallskip}
{}&
{}&
{(K)}&
{($10^7\,$cm$^{-3}$)}&
{($10^{13}\,$cm$^{-2}$)}&
{($10^{-5}$)}&
{($10^{-10}$)}\\
\hline
\noalign{\smallskip}
MMS6&t&    50&  1.0&3.0&0.7&0.3\\
MMS9&t& $<30$&  0.5&1.5&0.6&1.1\\

FIR4&n&    70&$>10$&30&&6.1\\
FIR4&b&    80&   5& 95&&19\\
FIR4&t&      &     &125&37&25\\

NGC2071&t&    60&  0.5&30&5.6&5.1\\
\\

PDR1&t&$>100$&  5.0&6.0&0.6&5.8\\
\hline
\noalign{\smallskip}
\end{tabular}
\label{t_meth}
\begin{list}{}{}
\item[$^{\mathrm{a}}$]
The letter designations refer to the velocity components with
narrow (n), broad (b), and total (t).
\end{list}
\end{table}

%% file: 3536_t8.tex
\begin{table}
\caption{Abundances Discerned from HCN, HNC, and CN}
\begin{tabular}{lccccccc}
\hline
\noalign{\smallskip}
{Source}&
{[HCN]/[HNC]}&
{$N$(HCN)}&
{$N$(CN)}&
{[HCN]/[CN]}&
{[HCN]/[H$_2$]}&
{[CN(total)]/[H$_2$]}\\
\noalign{\smallskip}
{}&
{}&
{($10^{14}\,$cm$^{-2}$)}&
{($10^{14}\,$cm$^{-2}$)}&
{($10^{0}$)}&
{($10^{-10}$)}&
{($10^{-10}$)}\\
\hline
\noalign{\smallskip}
MMS6&6.3&    0.6&$<$2.0&$>$0.3&0.6&$<$2.4\\
MMS9&5.4&    1.3&   2.8&   0.5&9.3&32.\\

FIR4&11.&    8.0&   6.0&   1.3&16.&30.\\
NGC2071&6.0& 6.4&   19.&   0.3&10.&41.\\
\\

PDR1&33.&    4.0&   2.0&   2.0&40.&72.\\
\hline
\noalign{\smallskip}
\end{tabular}
\label{t_hcn}
\end{table}

%% file: 3536_t9.tex
\begin{table}
\caption{Physical Conditions and Abundances Discerned from CS and SO}
\begin{tabular}{lccccccc}
\hline
\noalign{\smallskip}
{Source}&
{$N$(CS)}&
{[CS]/[H$_2$]}&
{$T_{\rm CS}$}&
{$N$(SO)}&
{[SO]/[H$_2$]}&
{[SO]/[CS]}\\
\noalign{\smallskip}
{}&
{($10^{13}\,$cm$^{-2}$)}&
{($10^{-10}$)}&
{(K)}&
{($10^{14}\,$cm$^{-2}$)}&
{($10^{-10}$)}&
{($10^{0}$)}\\
\hline
\noalign{\smallskip}
MMS6&3.5&     0.3&$>$50&$<$2.6&$<$2.5&$<$7.4\\
MMS9&1.5&     1.1&     &      &      &\\

FIR4&14.&     2.9&$>$100&$<$4.0&$<$8.2&$<$2.9\\
NGC2071&19.&  3.1&$>$ 50&	&\\
\\

PDR1&29.&     29.&$>$ 50&   0.8&  8.0&0.3\\
\hline
\noalign{\smallskip}
\end{tabular}
\label{t_cs}
\end{table}

%% file: 3536_cap.tex
\begin{figure}
\includegraphics[width=15cm]{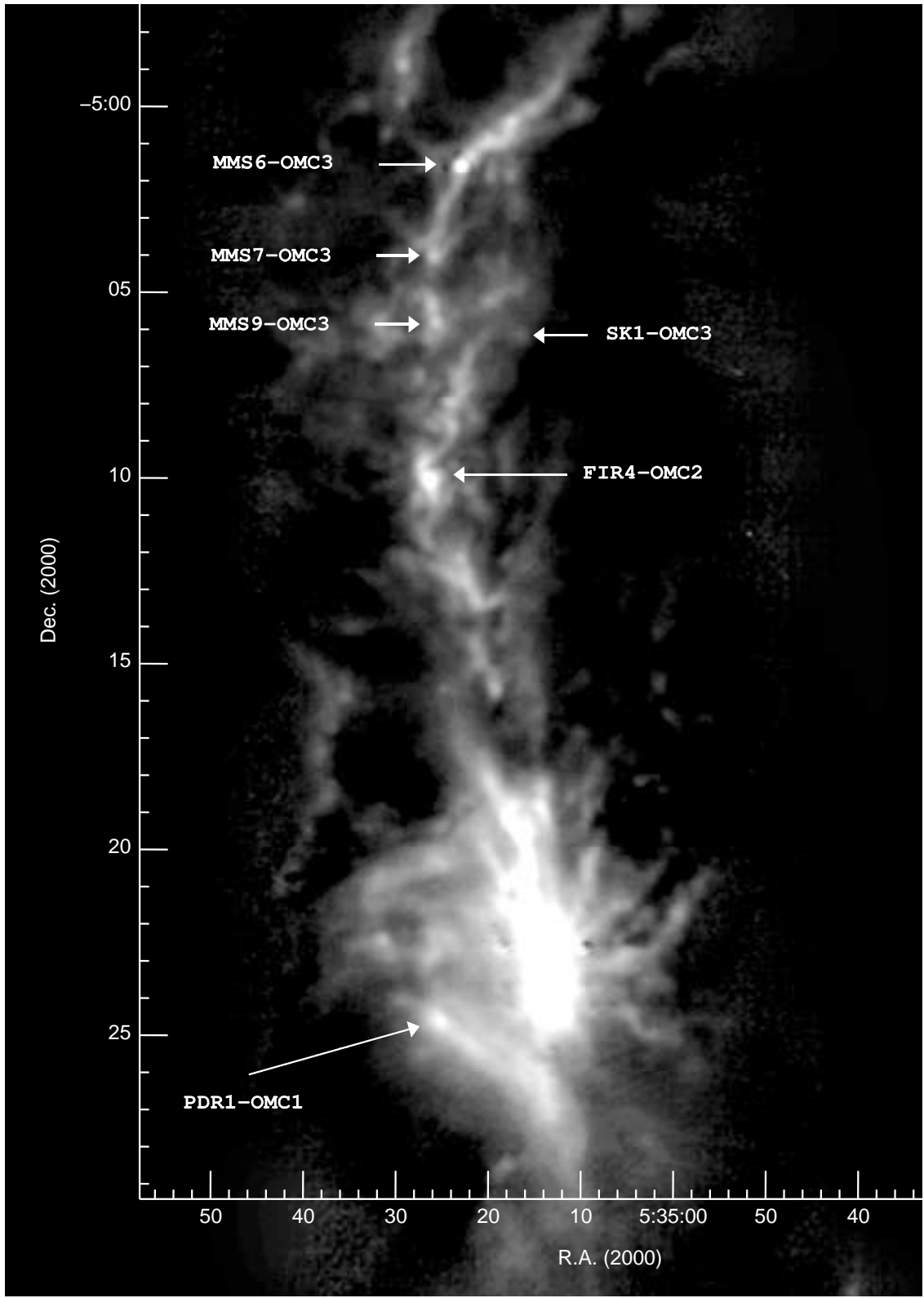}
\caption[]{
The Integral Shaped Filament (ISF) at the north end of the Orion A molecular cloud, observed
in dust continuum emission at 850 microns. The grey-scale ranges from 0 to 5 Jys with a
logarithmic stretch. The positions of the six sources located in the ISF are indicated
by arrows pointing at the peak continuum flux locations. Details of the sources are found
in Table \ref{t_prop}
\label{f_orion} }
\end{figure}
\clearpage

\begin{figure}
\includegraphics[width=15cm]{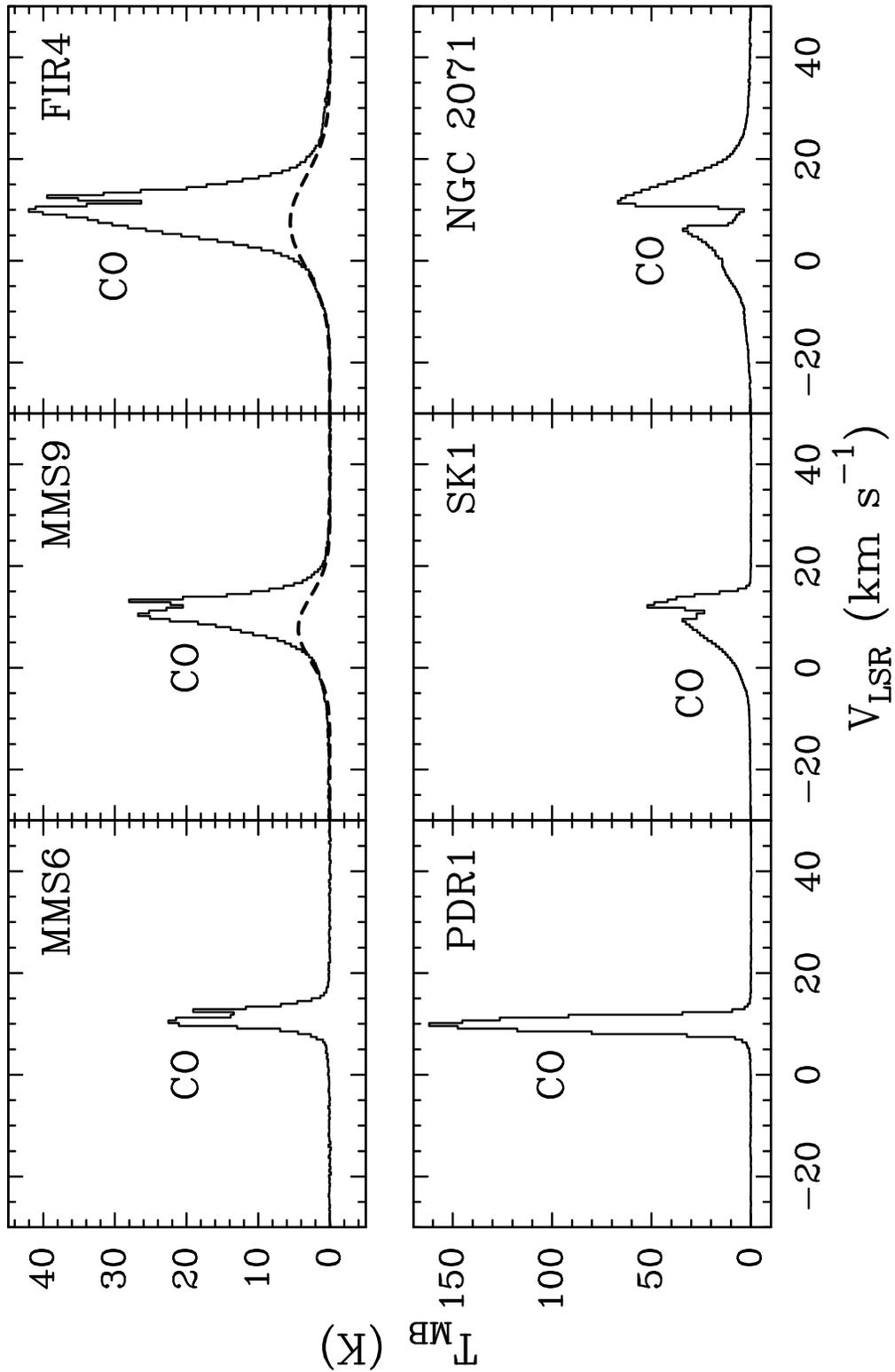}
\caption[]{
The CO $J=3$--$2$ line profiles. Note the self-absorption in all sources except PDR1,
and the extended line wings (dashed lines) in all sources except MMS6, and PDR1.
\label{f_cop} }
\end{figure}
\clearpage

\begin{figure}
\includegraphics[width=15cm]{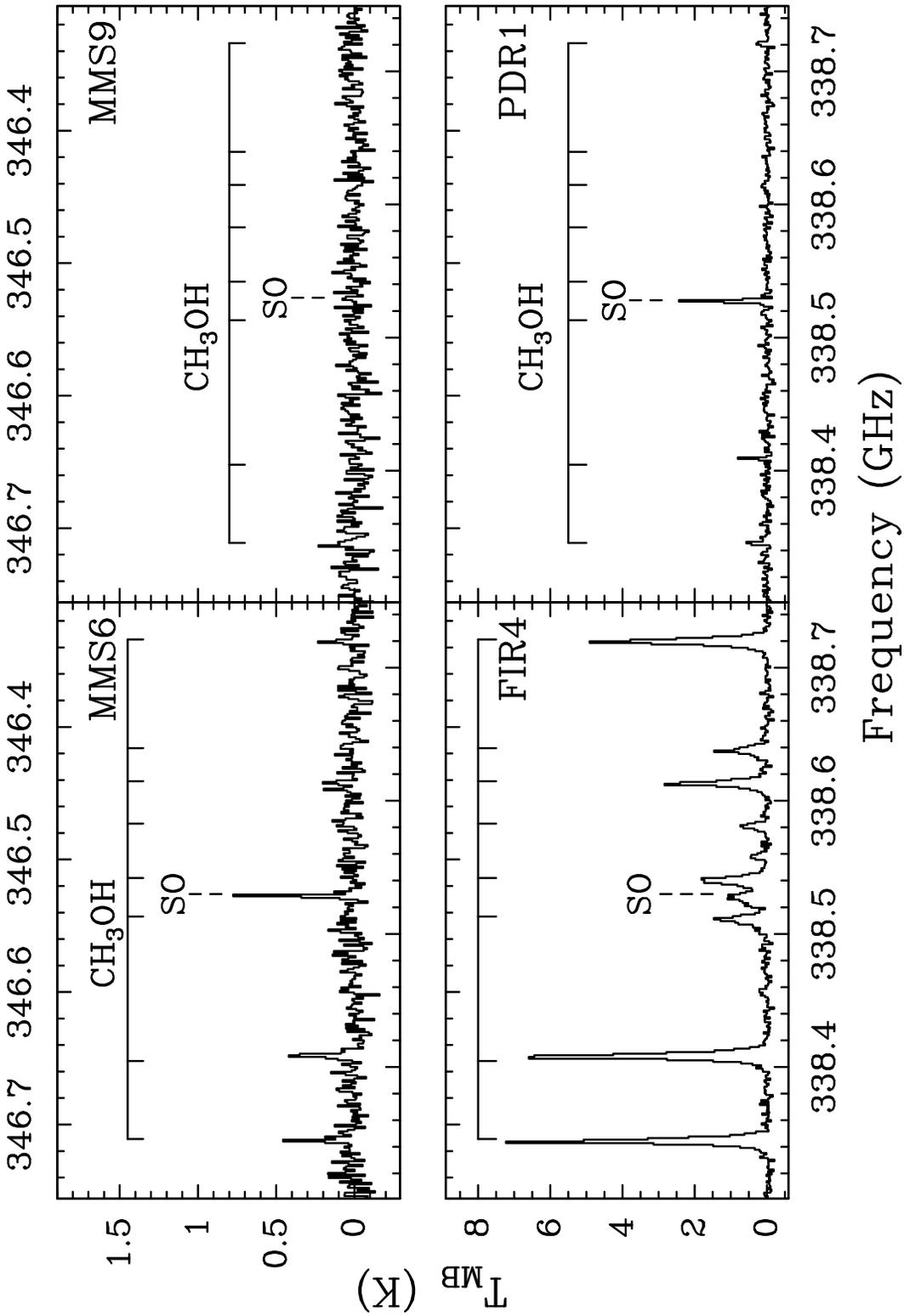}
\caption[]{
A selection of methanol lines observable with a single spectral setting, 
along with an SO $N_J$=$8_9\rightarrow7_8$ line.
\label{f_ch3ohp} }
\end{figure}
\clearpage

\begin{figure}
\includegraphics[width=15cm]{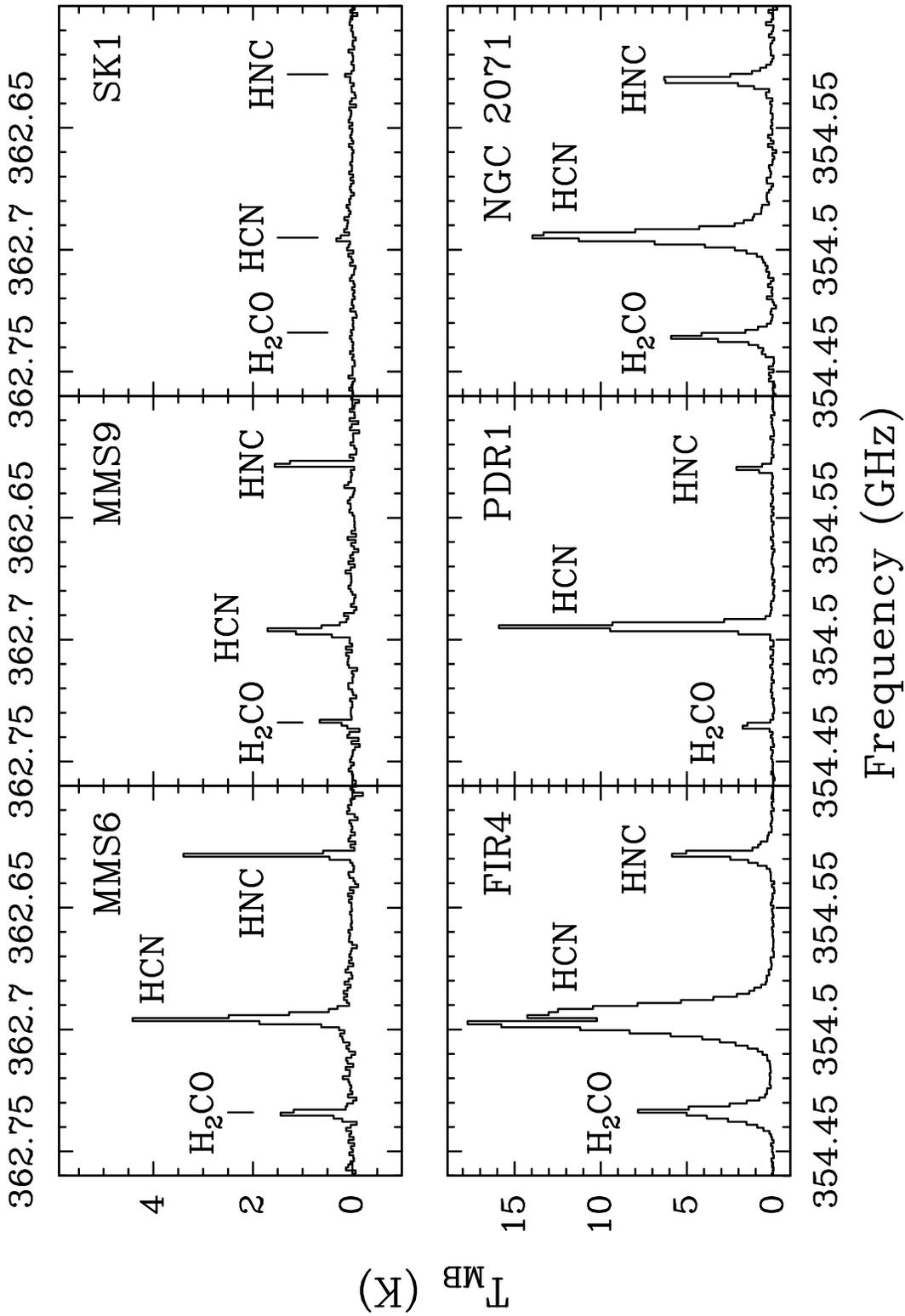}
\caption[]{
The HCN and HNC $J=4$--$3$ transitions. 
\label{f_hcnp} }
\end{figure}
\clearpage

\begin{figure}
\includegraphics[width=15cm]{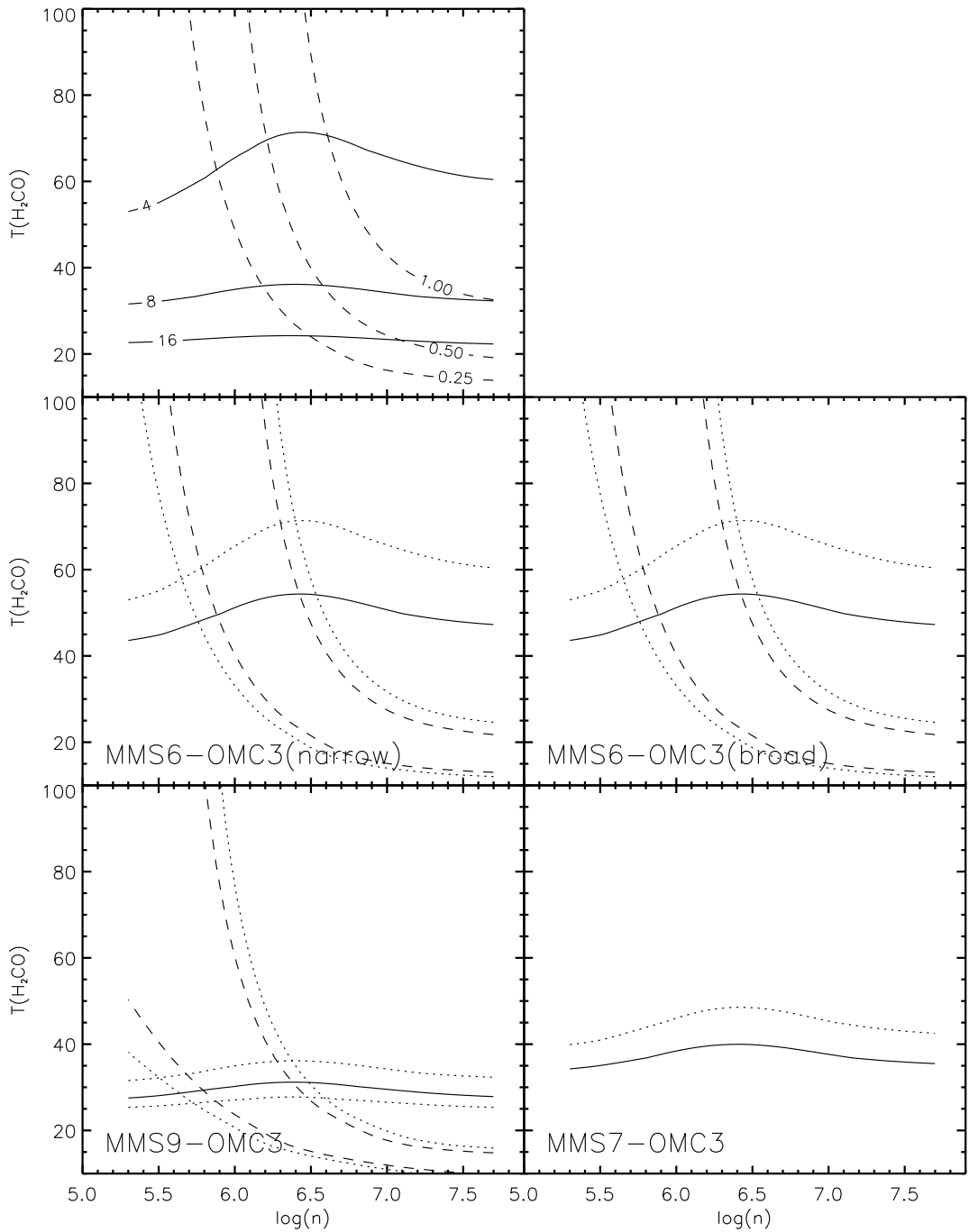}
\caption[]{
Determination of the equilibrium physical properties of molecular gas from
formaldehyde line ratios. The top left panel presents the expected line
ratio for the $3_{03} - 2_{02}$ and $3_{22} - 2_{21}$ transitions as
functions of temperature and density (solid contours) against the expected
line ratio for the $5_{05} - 4_{04}$ versus the $3_{03} - 2_{02}$ transitions
(dashed contours). The remaining panels show the range of physical properties
obtainable for the sources in the present study, with the solid lines
denoting the $3_{03} - 2_{02}$ versus $3_{22} - 2_{21}$ ratio and the
dashed lines providing the location of the $5_{05} - 4_{04}$ versus 
$3_{03} - 2_{02}$ ratio. Two sets of dashed lines are shown to account for
possible beam dilution (see text). Also, the dotted lines provide an
indication of the range of uncertainty on the calculations.
\label{f_h2co} }
\end{figure}
\clearpage

\begin{figure}
\includegraphics[width=15cm]{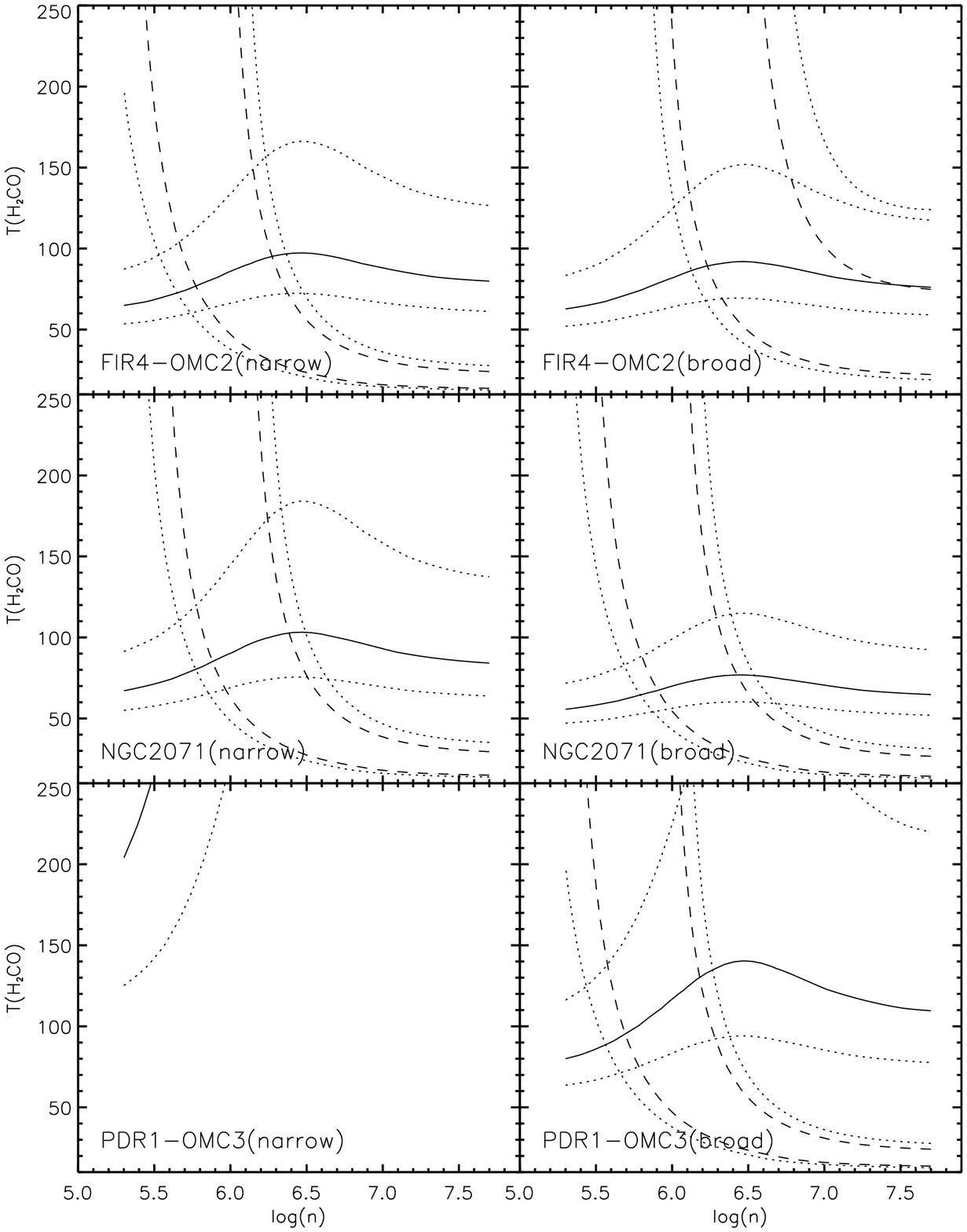}
\end{figure}
\clearpage

\begin{figure}
\includegraphics[width=9cm]{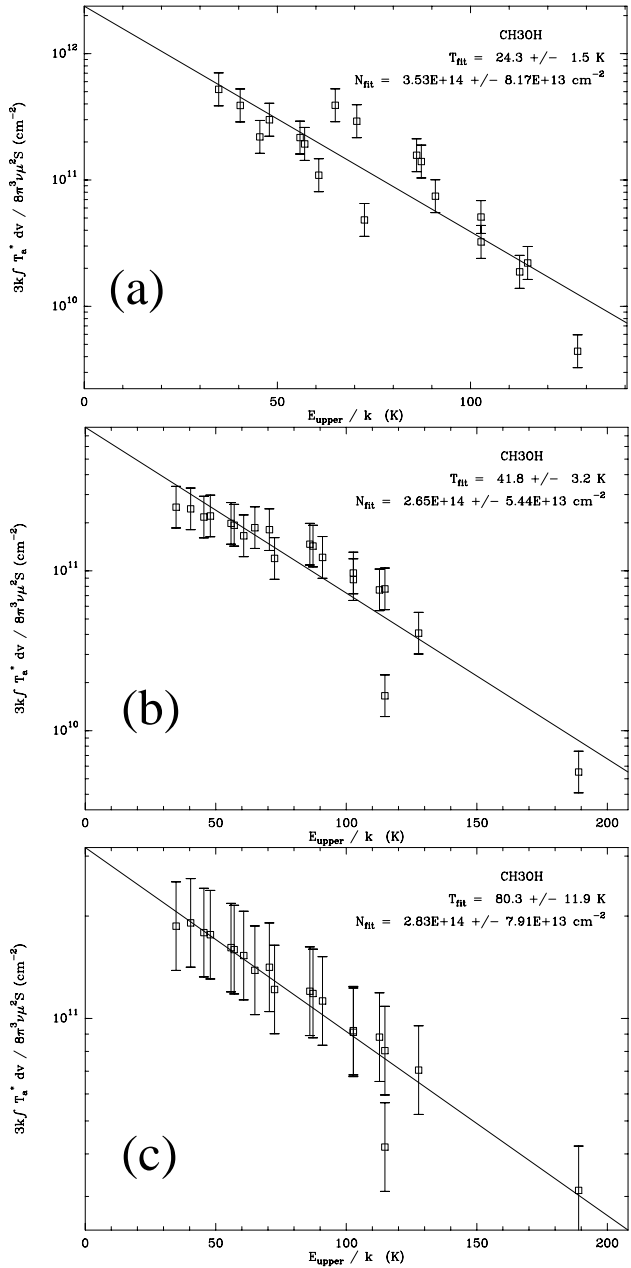}
\caption[]{
Rotation diagrams for methanol. 
The input data are taken from the result 
of RADEX equilibrium calculations at a temperature of 100\,K.
{\bf(a)} Calculation with $n = 10^7\,$cm$^{-3}$.
The slope of the data at high energies is significantly
affected by the sub-thermal excitation, and the derived temperature from
the rotation diagram does not correspond to the physical condition.
The low energy lines are much closer to being thermalized, however,
and the column density derived is close to the input column density of 
$10^{14}\,$cm$^{-3}$.
{\bf(b)} Calculation with $n = 10^8\,$cm$^{-3}$. The slope of the data at high 
energies is still affected by the sub-thermal excitation, and the derived 
temperature from the rotation diagram continues to underestimate the physical 
condition.
{\bf(c)} Calculation with $n = 10^9\,$cm$^{-3}$. 
At this density the rotation diagram produces a reasonable fit to the input 
physical conditions.
\label{f_ch3oh} }
\end{figure}
\clearpage

\begin{figure}
\includegraphics[width=15cm]{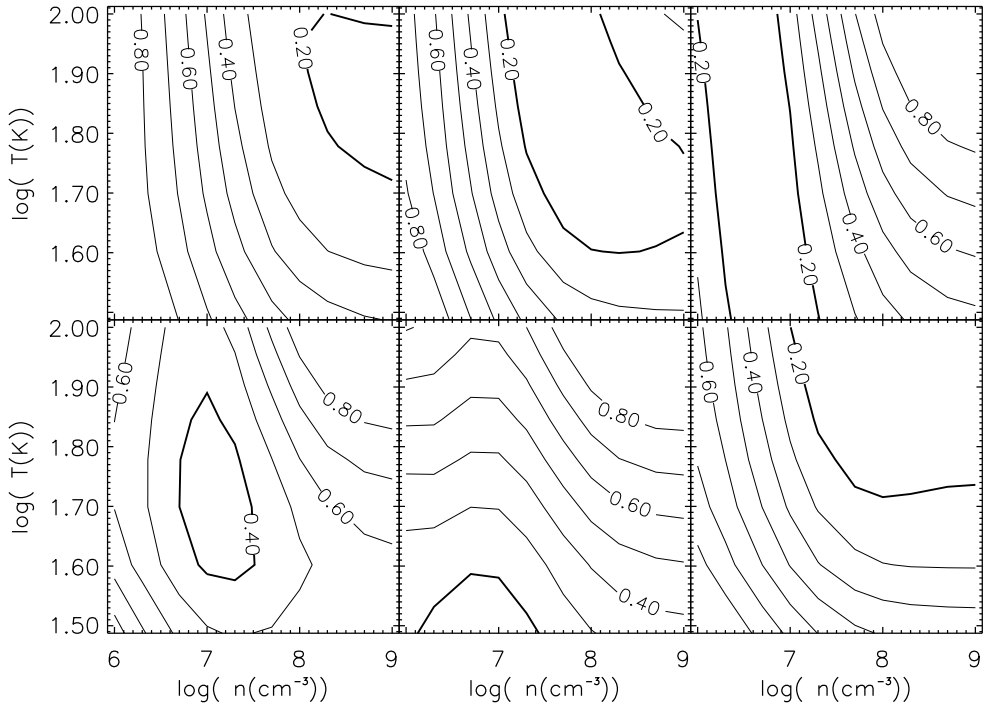}
\caption[]{
Chi-squared confidence fits to the physical parameters of sources using the observed
methanol lines and the radiative transfer code RADEX.
The sources are (Top Left) FIR4 narrow component, (Top Center) FIR4 broad component,
(Top Right) NGC2071, (Bottom Left) MMS6, (Bottom Center) MMS9, (Bottom Right) PDR1. 
Note that there is often a strong degeneracy between the best fit temperature
and density due to the lack of thermalization of the higher energy transitions.
\label{f_chi2} }
\end{figure}
\clearpage

\begin{figure}
\includegraphics[width=15cm]{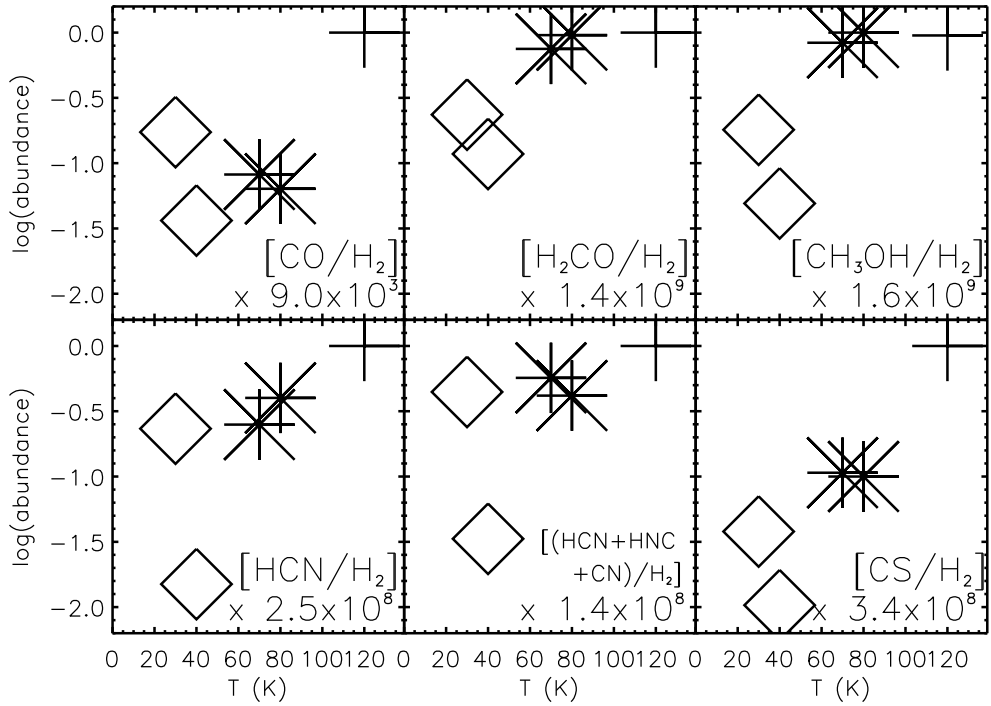}
\caption[]{
{
The range in abundances of the various molecular species investigated against
T(H$_2$CO). The abundances have been derived assuming a constant temperature
and density envelope. Diamonds represent sub-millimeter sources, stars 
represent infrared sources, and the cross represents the PDR, with symbol size 
approximating the order unity uncertainty expected in the measurements. 
For all protostellar sources the CO abundance is depleted by about an
order of magnitude. For all species the infrared sources show a remarkable
similarity in abundances, however, in the case of methanol the very broad
$>10\,$km$\,$s$^{-1}$ component from FIR4 has not been included. 
For most molecules the sub-millimeter sources have lower abundances while the
PDR has the highest abundance.
}

\label{f_abund} }
\end{figure}

%% file: 3536_tA1.tex
\begin{table}
\caption{Fractional population of the 3rd CO energy level for different 
physical conditions}
\begin{tabular}{lccc}
\hline
\noalign{\smallskip}
{}&
{\ \ \ \ \ \ \ \ \ \ \ \ \ \ \ \ \ \  Density (cm$^{-3}$)}&&\\
\noalign{\smallskip}
{Temperature (K)}&
{$10^3$}&
{$10^4$}&
{$10^5$}\\
\hline
\noalign{\smallskip}
10& 2.51e-3& 2.23e-2& 5.51e-2\\
20& 1.23e-2& 9.02e-2& 1.68e-1\\
30& 2.21e-2& 1.41e-1& 2.13e-1\\
50& 3.66e-2& 1.95e-1& 2.24e-1\\
100& 5.61e-2& 2.38e-1& 1.91e-1\\
500& 8.44e-2& 2.53e-1& 1.13e-1\\
\hline
\noalign{\smallskip}
\end{tabular}
\label{t_frac}
\end{table}

%% file: 3536_tA2.tex
\begin{table}
\caption{Ratio of CO line strength to 850 $\mu$m continuum with SCUBA for
 different physical conditions, assuming [CO]/[H$_2$] = $10^{-4}$ and $\kappa_
 {850} = 0.02\,$cm${^{2}}\,$g$^{-1}$.}
\begin{tabular}{lccc}
\hline
\noalign{\smallskip}
{}&
{\ \ \ \ \ \ \ \ \ \ \ \ \ \ \ \ \ \  Density (cm$^{-3}$)}&&\\
\noalign{\smallskip}
{Temperature (K)}&
{$10^3$}&
{$10^4$}&
{$10^5$}\\
\hline
\noalign{\smallskip}
 10& 0.4& 3.3& 8.2\\
 20& 0.8& 4.0& 7.5\\
 30& 0.6& 3.6& 5.4\\
 50& 0.5& 2.6& 3.0\\
100& 0.4& 1.5& 1.2\\
500& 0.1& 0.3& 0.1\\
\hline
\noalign{\smallskip}
\end{tabular}
\label{t_ratio}
\end{table}